\documentclass[10pt,journal,compsoc]{IEEEtran}
\ifCLASSOPTIONcompsoc
  \usepackage[nocompress]{cite}
\else
  \usepackage{cite}
\fi
\ifCLASSINFOpdf
\else
\fi
\usepackage{amsfonts}
\usepackage{bm}
\usepackage{amsmath}

\usepackage{amssymb,stmaryrd}
\usepackage{multirow} 
\usepackage{algpseudocode}
\usepackage{algorithm}
\usepackage{subfigure}
\usepackage{footmisc}
\usepackage{graphicx}
\usepackage{booktabs}
\usepackage{color}

\makeatletter 
\newcommand\figcaption{\def\@captype{figure}\caption} 
\newcommand\tabcaption{\def\@captype{table}\caption} 
\makeatother

\algnewcommand\algorithmicforeach{\textbf{for each}}
\algdef{S}[FOR]{ForEach}[1]{\algorithmicforeach\ #1\ \algorithmicdo}

\newcommand{\jy}[1]{\textcolor{black}{#1}}
\newcommand{\dk}[1]{\textcolor{black}{#1}}

\begin{document}
%
\title{Link Prediction with Contextualized Self-Supervision}
%
%
%
%

\author{Daokun Zhang,
        Jie Yin,~\IEEEmembership{Member,~IEEE}
        and Philip S. Yu,~\IEEEmembership{Life~Fellow,~IEEE}
\IEEEcompsocitemizethanks{\IEEEcompsocthanksitem Daokun Zhang is with the Department of Data Science \& AI, Faculty of Information Technology, Monash University, Australia,\protect\\
E-mail: daokun.zhang@monash.edu.
\IEEEcompsocthanksitem Jie Yin is with the Discipline of Business Analytics, The University of Sydney, Australia.
\protect\\
Email: jie.yin@sydney.edu.au.
\IEEEcompsocthanksitem Philip S. Yu is with Department of Computer Science, University of Illinois at Chicago, USA.
\protect\\
Email: psyu@uic.edu.
}
\thanks{Manuscript received April 19, 2005; revised August 26, 2015.}}

%
%

\markboth{Journal of \LaTeX\ Class Files,~Vol.~14, No.~8, August~2015}%
{Shell \MakeLowercase{\textit{et al.}}: Bare Demo of IEEEtran.cls for Computer Society Journals}
%



\IEEEtitleabstractindextext{%
\begin{abstract}
Link prediction aims to infer \dk{the link existence} between \dk{pairs of nodes} in networks/graphs. Despite their wide application, the success of traditional link prediction algorithms is hindered by three major challenges---\textit{link sparsity}, \textit{node attribute noise} and \jy{\textit{dynamic changes}}---that are faced by \dk{many} real-world networks. To \jy{address} these challenges, we propose a \underline{C}ontextualized \underline{S}elf-\underline{S}upervised \underline{L}earning (CSSL) framework that fully exploits structural context prediction for link prediction. The proposed CSSL framework learns a link encoder to \dk{infer the link existence probability from} paired node embeddings, which are constructed via a transformation on node attributes. To generate informative node embeddings for link prediction, structural context prediction is leveraged as a self-supervised learning task to boost the link prediction performance. Two types of structural context are investigated, \textit{i.e.}, context nodes collected from random walks \textit{vs.} context subgraphs. The CSSL framework can be trained in an end-to-end manner, with the learning of \dk{model parameters} supervised by \dk{both the} link prediction and self-supervised learning tasks. The proposed CSSL is a generic and flexible framework in the sense that it can handle both attributed and non-attributed networks, and \dk{operate under} both transductive and inductive link prediction settings. Extensive experiments and ablation studies on seven real-world benchmark \dk{networks} demonstrate the superior performance of the proposed self-supervision based link prediction algorithm over state-of-the-art baselines, on different types of networks under both transductive and inductive settings. The proposed CSSL also yields competitive performance in terms of its robustness to node attribute noise and scalability over large-scale networks.
\end{abstract}

}

\maketitle

\IEEEdisplaynontitleabstractindextext

%
\IEEEpeerreviewmaketitle

\IEEEraisesectionheading{\section{Introduction}\label{sec:introduction}}

Link prediction is an increasingly important task on graph-structured data with broad applications, such as friend recommendation~\cite{xie2010potential}, knowledge graph completion~\cite{nathani2019learning}, entity resolution~\cite{fu2019end}, and targeted advertising~\cite{fan2019graph}, \textit{etc}. A straightforward solution to link prediction is to calculate some heuristic metrics, such as Common Neighbors~\cite{newman2001clustering}, Adamic-Adar Index~\cite{adamic2003friends} and Katz Index~\cite{katz1953new}, \textit{etc}., to infer the link existence. Recently, learning based \dk{link prediction} methods ~\cite{kipf2016variational,zhang2018link} have been proved more effective, as they are able to \jy{leverage not only richer structure semantics but also essential node attributes of networks}.

Nevertheless, the success of learning based link prediction methods relies heavily on the quality and static property of \dk{targeted} networks. However, \jy{performing link prediction on real-world networks} are \jy{usually confronted with} three key challenges as follows:
\begin{itemize}
	\item \textbf{Link sparsity.} Due to access restrictions or privacy concerns, most of real-world networks are sparse in the sense that the number of unconnected node pairs in an observed network grows quadratically as the number of links grows linearly~\cite{ghasemian2020stacking}. \jy{Akin to} supervised learning \dk{for non-relational data}, link sparsity incurs insufficient supervision for link prediction, resulting in poor performance. 
	\item \textbf{Node attribute noise.} Real-world networks are often corrupted with node attribute noise, for example, inaccurate user profiles in social networks.  
	This compromises the link prediction accuracy, particularly for learning based methods that \jy{highly} rely on node attributes, \dk{for which} incorrect mappings from node attributes to link existence could be learned.
	\item \jy{\textbf{Dynamic changes.}} In many cases, networks are not static but dynamically changing with new nodes joining constantly, where no or only few links are observed. For these out-of-sample nodes, the absence or sparsity of their neighborhood structure makes it difficult for many learning based methods to perform link prediction accurately. 
\end{itemize}


\begin{figure*}[t]
	\centering
	\includegraphics[width=1.835\columnwidth]{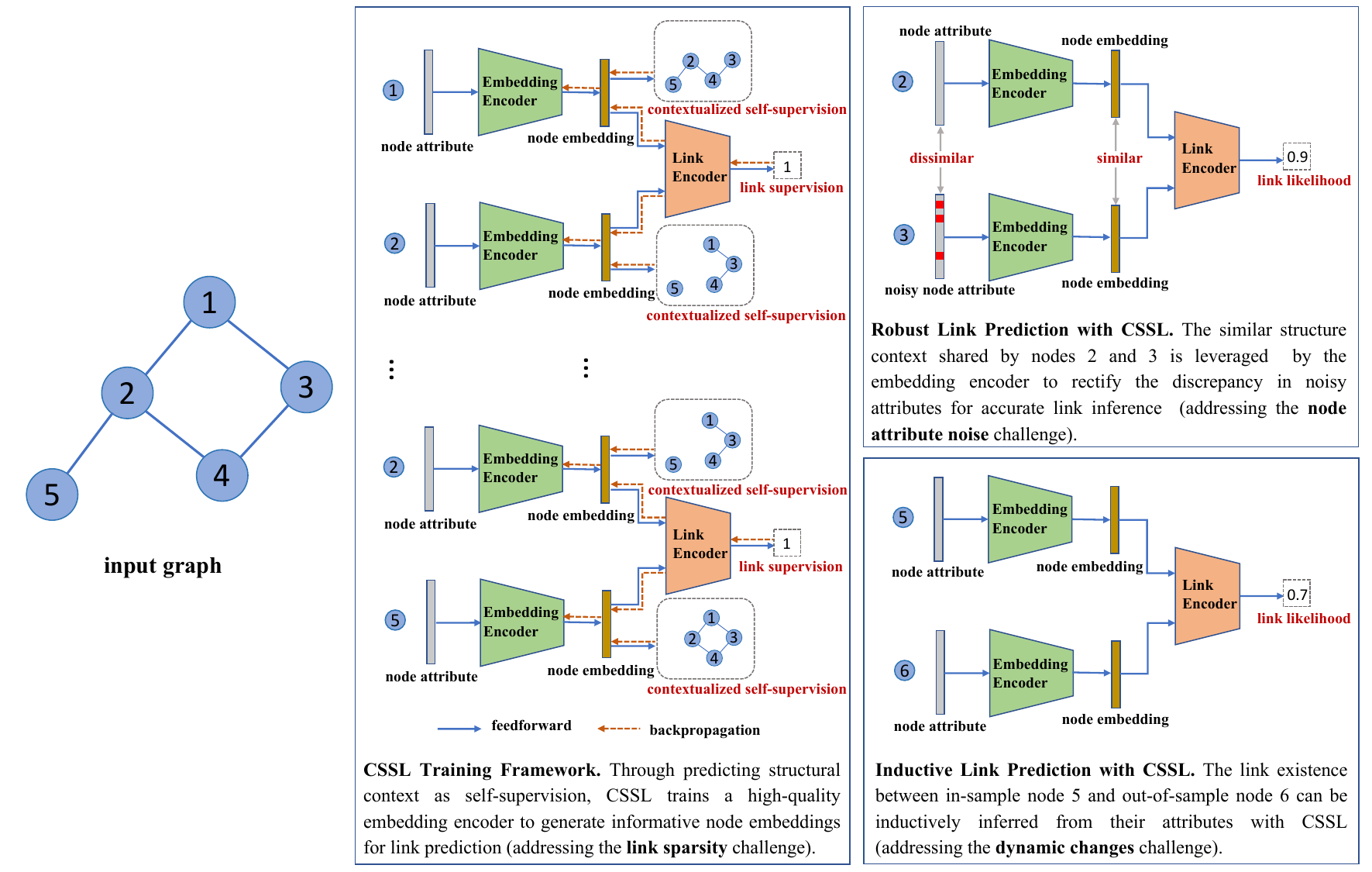} 
	\caption{\jy{An illustration of the proposed CSSL framework with the key idea to address the three key challenges confronted by link prediction: \textbf{link sparsity}, \textbf{node attribute noise} and \textbf{dynamic changes}.}}
	\label{CSSL_diagram}
\end{figure*}

\dk{As a paradigm of} semi-supervised learning, self-supervised learning (SSL)~\cite{kolesnikov2019revisiting} has been leveraged as an effective strategy in computer vision to favor the supervised learning task through learning high-quality \dk{data} representations from the unlabeled data. This has motivated us to exploit self-supervision to tackle the \textbf{link sparsity} challenge for link prediction. Although recent attempts have been made to exploit self-supervision on graphs for node classification via node-oriented pretext tasks such as node attribute prediction~\cite{you2020does}, they are not designed in an effective way to capture \dk{structural patterns that are essential} for link prediction. To address the \textbf{node attribute noise} challenge, we propose to bridge self-supervised learning with structural context modeling to reinforce link prediction, as structural context has been proved beneficial to node representation learning~\cite{zhang2017user} and attribute selection~\cite{tang2012feature}. To effectively predict the links of out-of-sample nodes (addressing the \textbf{dynamic changes} challenge), we design the main link prediction task as an end-to-end learning framework that \dk{leverages} attributes of paired nodes to infer their link existence probability. With the structural context modeling based self-supervised learning, we can learn the link prediction model as an accurate mapping from node attributes to link existence, with \dk{stronger} generalization ability \dk{for handling} out-of-sample nodes. \jy{As such}, the links of out-of-sample nodes can be accurately predicted by applying the trained model to the attributes of out-of-sample nodes, which effectively overcomes the negative impact caused by the absence or sparsity of their neighborhood structure.


Based on the above \dk{insights}, we propose a new contextualized self-supervised learning (CSSL) framework for link prediction. The proposed CSSL framework exploits the power of structural context prediction based self-supervision to enable context-aware link prediction. Specifically, CSSL formulates a general learning framework that \jy{learns a link encoder} to infer the link existence probability from paired node embeddings constructed from node attributes. The \dk{link encoder predicts the link existence probability through constructing edge embeddings from node embeddings, which} effectively capture the attribute difference/similarity between connected nodes. \jy{This new formulation enables us to exploit} the power of the homophily property~\cite{mcpherson2001birds} for link prediction---the fact that connected node pairs usually have similar attributes. The structural context prediction is modeled as a self-supervised learning task to boost the link prediction performance. We investigate two types of contextualized self-supervised learning tasks---context node prediction and context subgraph prediction, \dk{where context nodes and context subgraphs} are sampled through short random walks. The proposed CSSL framework can be trained efficiently with stochastic gradient descent by sampling a minibatch of edges at each iteration and thus have a time complexity linear to the number of edges. As CSSL learns a mapping function from node attributes to the link existence, it \jy{has the ability to inductively} predict the links for out-of-sample nodes, by applying the learned mapping function to the attributes of out-of-sample nodes. \dk{Besides, CSSL can be naturally extended to non-attributed networks through using one-hot encodings as node attributes.}

\jy{Fig.~\ref{CSSL_diagram} provides an illustration to elaborate how the proposed CSSL framework addresses the three aforementioned link prediction challenges. To handle the \textbf{link sparsity} challenge, 
CSSL trains a probabilistic link prediction model more effectively via 
contextualized self-supervision. This is accomplished by leveraging 
the auxiliary link existence
evidence conveyed in structural context,
which can help train a high-quality embedding encoder to generate informative node embeddings for link prediction.
To tackle the \textbf{attribute noise} challenge, CSSL makes the best of structural context to rectify the discrepancy in noisy node attributes for robust link inference. In case of predicting the link existence between nodes 2 and 3, where attributes of node 3 are corrupted with noise, the well-trained embedding encoder allows CSSL to generate similar node embeddings from their common structural context, thereby alleviating the adverse impact of attribute noise on link prediction. To address the \textbf{dynamic changes} challenge, CSSL can inductively infer the link existence between out-of-sample nodes (\textit{e.g.}, node 6) and in-sample nodes (\textit{e.g.}, node 5) from their attributes, without the dependence on their neighborhood structure.}


We conduct extensive experiments on \dk{seven} real-world \dk{attributed} networks, which show that the proposed CSSL framework outperforms the state-of-the-art methods by large margins, under both transductive and inductive settings. This proves the advantages of CSSL in making the best of network structure and extracting useful information from noisy node attributes towards high-quality link prediction. \jy{In addition, we verify its scalability to large-scale networks and its significant advantages over competitive baselines for link prediction on non-attributed networks.}

The contribution of this paper is threefold:
\begin{itemize}
	\item We analyze the negative impact of link sparsity, node attribute noise and dynamic changes on link prediction, which motivates the design of structural context prediction based self-supervised learning task for link prediction. 
	\item We propose the first end-to-end contextualized self-supervision based link prediction framework that is inductive with the ability to predict the links for both in-sample and out-of-sample nodes and has the flexibility to handle both attributed and non-attributed networks. 
	\item Comprehensive experiments and ablation studies show the superiority of our approach over competitive baselines under both transductive and inductive settings, as well as its robustness to node attribute noise and scalablity to large-scale networks. 
\end{itemize}

The rest of this paper is organized as follows. In Section 2, we review related work on link prediction. The link prediction problem is then formally defined in Section 3. In Section 4, we elaborate our proposed self-supervision based link prediction framework. Extensive experimental results are then reported in Section 5 to evaluate the proposed algorithm. Finally, in Section 6, we conclude this paper.

\section{Related Work}
This section briefly reviews two branches of related work on link prediction and self-supervised learning.

\subsection{Link Prediction on Graphs}
Traditional link prediction methods typically infer link existence by calculating some heuristic metrics on graphs, such as Common Neighbors~\cite{newman2001clustering}, Jaccard Index~\cite{jaccard2001etude}, Preferential Attachment Index~\cite{barabasi1999emergence}, Adamic-Adar Index~\cite{adamic2003friends}, Katz Index~\cite{katz1953new}, 
Rooted PageRank~\cite{page1999pagerank}, and SimRank~\cite{jeh2002simrank}. However, heuristic methods simply define rigid and one-sided structural measures to infer the link existence between node pairs, failing to capture the richer structural context and the essential information conveyed by node attributes. Recently, learning based methods have been proposed to advance link prediction under two categories. 

\subsubsection{Network embedding based methods}
Network embedding based methods first learn node embeddings via network embedding techniques, and then train a classifier to perform link prediction by taking edge embeddings aggregated from node embeddings as features. A series of network embedding algorithms can be used to learn node embeddings. DeepWalk~\cite{perozzi2014deepwalk} and Node2Vec~\cite{grover2016node2vec} learn node embeddings by using them to predict random walk context nodes. LINE~\cite{tang2015line} learns node embeddings by modeling the similarity between connected nodes and the similarity between nodes sharing common neighbors. SDNE~\cite{wang2016structural} and DNGR~\cite{cao2016deep} learn deep node \dk{embeddings} with a deep auto-encoder neural network. GraphSAGE~\cite{hamilton2017inductive} and Graph Convolutional Networks (GCNs)~\cite{kipf2016semi} learn node \dk{embeddings} through aggregating the attributes\dk{/embeddings} of neighboring nodes. CANE~\cite{tu2017cane} learns context-aware node \dk{embeddings} by capturing the attribute attention between connected nodes. MVC-DNE~\cite{yang2017properties} fuses network structure and node attributes into node \dk{embeddings} with a deep multi-view auto-encoder. Attri2Vec~\cite{zhang2019attributed} learns node embeddings by projecting node attributes into a structure-aware subspace. Most of the existing network embedding algorithms are agnostic to the downstream link prediction task in a way that specific contextual information required for link prediction is not well captured. Moreover, these methods lack the ability to accurately infer embeddings of out-of-sample nodes \dk{that have} few or no links for performing link prediction inductively. 

\subsubsection{Link modeling based methods}
Link modeling based methods directly learn a model to infer link labels from the inputs of paired nodes with node attributes and/or node neighborhood structure. Various techniques are proposed to devise end-to-end link inference models. Fact~\cite{menon2011link} predicts the link existence between paired nodes from their attributes by using the bilinear regression model~\cite{ruben1998generalized}. NARM~\cite{zhao2002leveraging} designs a Bayesian probabilistic generative model to infer links from node attributes. \jy{However}, Fact and NARM use only node attributes to predict links. VGAE~\cite{kipf2016variational} reconstructs network links \dk{from} node embeddings obtained by graph convolution via Variational Auto-encoder (VAE)~\cite{rezende2014stochastic}. SEAL~\cite{zhang2018link} extracts local subgraphs around nodes and learns a function that maps the extracted subgraphs to link existence. SEAL is claimed to be able to learn heuristics that suit the current network, thus achieving better performance as compared to various heuristics. However, SEAL can only use node attributes as side information, but not in an integrated framework for link prediction. DEAL~\cite{hao2020inductive} predicts links by ensembling the link existence probabilities respectively estimated from node attribute-oriented embeddings and structure-oriented embeddings, where the two types of embeddings are aligned by maximizing their consistency. VGAE and SEAL require rich neighborhood structure to predict the links of out-of-sample nodes, \jy{whereas} Fact, NARM and DEAL predict links from only node attributes, warranting their inductive ability. 

In summary, all of these learning based link prediction methods rely \jy{merely} on a single link existence prediction task \jy{for model training. As a consequence,} they are ineffective with few training links and noisy node attributes. 
\subsection{Self-supervised Learning}

Self-supervised learning (SSL) is a promising learning paradigm for deep neural networks (DNNs) in the \jy{domains} of computer vision~\cite{Hendrycks2019UsingSL,Goyal2019ScalingAB} and natural language processing~\cite{devlin2019bert,lan2019albert}. \jy{The central theme of SSL is focused on defining} self-supervised pretext tasks to assist DNNs in learning more generalized and robust \dk{data} representations from the unlabeled data. The training strategies fall into two categories. 1) Pretraining: A DNN model is first pretrained with self-supervised pretext tasks and then finetuned with the target supervision task. 2) Joint training: The self-supervised tasks and the target supervision task are jointly trained with a multi-task learning objective.  

Recently, SSL has been explored to boost the performance of machine learning tasks on graph-structured data through defining various pretext tasks~\cite{liu2021graph,jin2020self}. Inspired by image inpainting in computer vision~\cite{yu2018generative}, node attribute reconstruction has been utilized to enhance the effectiveness of node embeddings learned by GCNs~\cite{you2020does,jin2020self,hu2019strategies,manessi2021graph}. On the other hand, node structural properties have also been leveraged to design pretext tasks, such as node degree~\cite{jin2020self}, structural distance between nodes~\cite{jin2020self}, graph partition affiliations of nodes~\cite{sun2020multi,you2020does}, node subgraph motifs~\cite{rong2020self}, and node centrality~\cite{hu2019pre}. The \jy{above-mentioned} self-supervised pretext tasks are designed with the pipeline of predicting attribute/structure properties from the learned node embeddings, which mainly favor the node classification task. Pioneered by DIG~\cite{velickovic2019deep}, another line of graph self-supervised learning focuses on maximizing the mutual information between node embeddings and the embeddings of their high-level graph summaries.  DIG~\cite{velickovic2019deep} and MVGRL~\cite{hassani2020contrastive} achieve this \jy{objective} by contrasting node \dk{embeddings} and the global graph \dk{representation}. Extensions are also explored to contrast node embeddings and the embeddings of their surrounding subgraphs~\cite{jiao2020sub,hu2019strategies,mavromatis2020graph,wang2021self,peng2020graph}. Moreover, the idea has been generalized to improve graph \jy{representation} learning by contrasting local subgraph embeddings and the global graph \jy{representation}~\cite{sun2019infograph,zhang2020motif,sun2021sugar}.  

To date, the current graph self-supervised learning algorithms are primarily designed to favor node-level and graph-level classification tasks. 
However, how to leverage self-supervision to facilitate link prediction is still under-explored. To fill this research gap, our work thoroughly analyses the roles of structural context in link-forming mechanisms on graphs and explicitly exploits structural context prediction as self-supervision to boost the link prediction performance.

\section{Problem Definition}
Assume we are given an attributed network $\mathcal{G}=(\mathcal{V},\mathcal{E},\bm{X})$, where $\mathcal{V}$ is the set of nodes, $\mathcal{E}$ is the set of edges, and $\bm{X}\in\mathbb{R}^{m\times|\mathcal{V}|}$ is node \dk{attribute} matrix with $m$ being the attribute dimension and the $i$-th column of $\bm{X}$, $\bm{x}_{i}\in\mathbb{R}^{m}$, being the attribute vector of the $i$-th node $v_i$ in $\mathcal{V}$. For the edge set $\mathcal{E}$, we have  $\mathcal{E}=\mathcal{E}_{tr}\cup\mathcal{E}_{te}$, where $\mathcal{E}_{tr}$ and $\mathcal{E}_{te}$ are the sets of training and test links, respectively. For \textit{transductive} link prediction, we assume that all \dk{nodes} in $\mathcal{V}$, connected by training and test links, are available, so we have the training graph as $\mathcal{G}_{tr}=(\mathcal{V},\mathcal{E}_{tr},\bm{X})$. In contrast, for \textit{inductive} link prediction, we assume that only nodes connected by training links are available, forming the set $\mathcal{V}_{tr}$, so we have the training graph as $\mathcal{G}_{tr}=(\mathcal{V}_{tr},\mathcal{E}_{tr},\bm{X}_{tr})$, where $\bm{X}_{tr}\in\mathbb{R}^{m\times|\mathcal{V}_{tr}|}$ \dk{stores} the attributes of nodes in $\mathcal{V}_{tr}$. 

Our objective is to learn a link prediction model on the training graph $\mathcal{G}_{tr}=(\mathcal{V},\mathcal{E}_{tr},\bm{X})$ or $(\mathcal{V}_{tr},\mathcal{E}_{tr},\bm{X}_{tr})$ by training the link prediction and contextualized self-supervised learning tasks jointly, with sparse training links \dk{in} $\mathcal{E}_{tr}$ and noisy node \dk{attributes} $\bm{X}$/$\bm{X}_{tr}$. The main link prediction task aims to train a function $f: \mathbb{R}^m \times\mathbb{R}^{m}\rightarrow[0,1]$, a mapping from $m$-dimensional attributes of paired nodes to the link existence probability between them, which can be used to predict the existence of unseen test links \dk{in} $\mathcal{E}_{te}$. By leveraging the contextualized self-supervision, the trained link prediction model is expected to have the following three properties: 1) the ability to capture the link existence evidence carried by links, node attributes and their interactions; 2) the robustness against link sparsity and node attribute noise; 3) the inductive ability to accurately predict the links of out-of-sample nodes.

\begin{figure}[t]
	\centering
	\includegraphics[width=0.9\columnwidth]{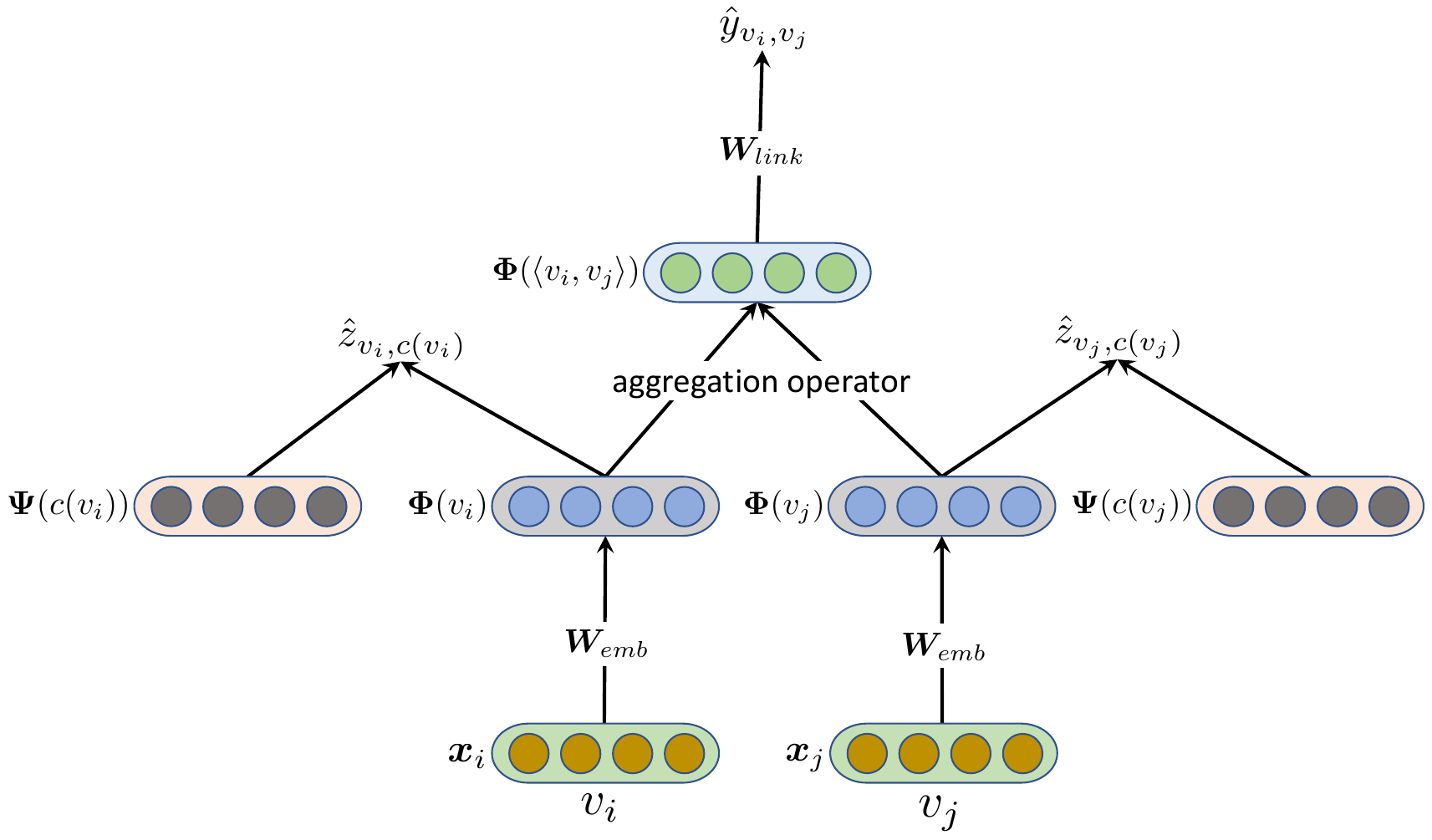} 
	\caption{The architecture of the proposed CSSL framework for link prediction. Node embeddings $\bm{\Phi}(v_i)$ and $\bm{\Phi}(v_j)$ are first constructed from $v_i$'s attribute $\bm{x}_i$ and $v_j$'s attribute $\bm{x}_j$, respectively. $\bm{\Phi}(v_i)$ and $\bm{\Phi}(v_j)$ are then aggregated to form edge embedding $\bm{\Phi}(\langle v_i,v_j \rangle)$, which is used to predict the link existence probability $\hat{y}_{v_i,v_j}$ between $v_i$ and $v_j$. The existence probability $\hat{z}_{v_i,c(v_i)}$ of $v_i$'s structural context $c(v_i)$, and the existence probability $\hat{z}_{v_j,c(v_j)}$ of $v_j$'s structure context $c(v_j)$ are respectively predicted from $\bm{\Phi}(v_i)$ together with the context embedding $\bm{\Psi}(c(v_i))$, and $\bm{\Phi}(v_j)$ together with the context embedding $\bm{\Psi}(c(v_j))$.}
	\label{CSSL_illustration}
\end{figure}

\section{CSSL for Link Prediction}

Existing learning based link prediction approaches~\cite{menon2011link,zhao2002leveraging,kipf2016variational,zhang2018link,hao2020inductive} mainly focus on minimizing the link classification loss on training \jy{links} \dk{using edge embeddings as features}, with the existing edges labeled with 1 and \dk{unseen} edges labeled with 0. The edge embeddings are constructed from node \dk{attributes}.
However, the existing methods rely heavily on \dk{a large quantity of} edge labels. On sparse networks, they tend to perform poorly without sufficient edge label supervision. With only a single type of supervision, they are also vulnerable to node attribute noise.

In this section, we present the proposed contextualized self-supervised learning (CSSL) framework for link prediction. In the proposed CSSL framework, node embeddings constructed from node attributes are aggregated into edge embeddings for link prediction. Meanwhile, node embeddings are used to predict the corresponding structural context. Through joint learning, the useful information in sparse links and noisy node attributes can be extracted and fully utilized to improve the link prediction performance. 

The architecture of the proposed CSSL framework is shown in Fig.~\ref{CSSL_illustration}. Given a node pair $\langle v_i, v_j\rangle$, we first transform their node attributes into a low-dimensional latent space to obtain their node embeddings. Using their node embeddings, we simultaneously perform two learning tasks. The first is a link prediction task supervised by the training links, which predicts the link existence probability between nodes $v_i$ and $v_j$ from the edge embedding aggregated from \jy{their} node embeddings. The second is to predict the occurrence probability of \dk{their} structural contexts. The structural context prediction is used in a self-supervised manner to boost the link prediction performance.

\subsection{Supervised Link Prediction Task}
Following traditional learning based algorithms~\cite{menon2011link,zhao2002leveraging}, we cast the link prediction problem as a binary classification task on edge embeddings constructed from \dk{the attributes} of paired nodes. With the carefully designed architecture, the constructed edge embeddings well capture node attribute difference/similarity, exerting the power of the homophily property~\cite{mcpherson2001birds} towards high-quality link prediction. The overall pipeline from node attributes to the link \jy{existence} likelihood prediction is end-to-end, which can be taken as a generalization of the existing learning based attribute-driven link prediction methods~\cite{menon2011link,zhao2002leveraging}.

For nodes $v_i$ and $v_j$, their node embeddings are respectively constructed from their attribute vectors $\bm{x}_{i}, \bm{x}_{j}\in \mathbb{R}^{m}$:
\begin{equation}
\begin{aligned}
\bm{\Phi}(v_i)=\sigma(\bm{W}_{emb}\;\bm{x}_{i}),\ \ \
\bm{\Phi}(v_j)=\sigma(\bm{W}_{emb}\;\bm{x}_{j}),
\end{aligned}
\label{node_emb}
\end{equation}
where $\bm{\Phi}(v_i)$ and $\bm{\Phi}(v_j)\in\mathbb{R}^{d}$ are the embeddings of node $v_i$ and node $v_j$ with $d$ dimensions, $\bm{W}_{emb}\in\mathbb{R}^{d\times m}$ is the weight matrix for constructing node embeddings from node attributes, and $\sigma(\cdot)$ is the sigmoid function. From node embeddings $\bm{\Phi}(v_i)$ and $\bm{\Phi}(v_j)$, we construct edge embedding $\bm{\Phi}(\langle v_i,v_j \rangle)\in\mathbb{R}^{d}$ for edge $\langle v_i,v_j \rangle $ with the following aggregation operators~\cite{grover2016node2vec}:
\begin{equation}
\begin{aligned}
&\bullet \text{Average:}\;\bm{\Phi}(\langle v_i,v_j \rangle)_r=(\bm{\Phi}(v_{i})_{r}+\bm{\Phi}(v_{j})_{r})/2,\\
&\bullet \text{Hadamard:}\;\bm{\Phi}(\langle v_i,v_j \rangle)_r=\bm{\Phi}(v_{i})_{r}\cdot\bm{\Phi}(v_{j})_{r},\\
&\bullet \text{Weighted-L1:}\;\bm{\Phi}(\langle v_i,v_j \rangle)_r=|\bm{\Phi}(v_{i})_{r}-\bm{\Phi}(v_{j})_{r}|,\\
&\bullet \text{Weighted-L2:}\;\bm{\Phi}(\langle v_i,v_j \rangle)_r=(\bm{\Phi}(v_{i})_{r}-\bm{\Phi}(v_{j})_{r})^{2},
\end{aligned}
\label{aggregation_operator}
\end{equation}where $\bm{\Phi}(\langle v_i,v_j \rangle)_r$, $\bm{\Phi}(v_{i})_{r}$ and $\bm{\Phi}(v_{j})_{r}$ indicate the $r$-th dimension of the embedding vectors $\bm{\Phi}(\langle v_i,v_j \rangle)$, $\bm{\Phi}(v_{i})$ and $\bm{\Phi}(v_{j})$, respectively.

On the constructed edge embedding $\bm{\Phi}(\langle v_i,v_j\rangle)$, the link existence probability $\hat{y}_{v_i,v_j}$ between nodes $v_i$ and $v_j$ is predicted as
\begin{equation}
\hat{y}_{v_i,v_j}=\sigma(\bm{W}_{link}\;\bm{\Phi}(\langle v_i,v_j\rangle)),
\label{link_inference}
\end{equation}where $\bm{W}_{link} \in \mathbb{R}^{1\times d}$ is the weight matrix for predicting the link existence probability. The link prediction task aims to minimize the following loss:
\begin{equation}
\mathcal{L}_{v_i,v_j}^{link}=\ell(y_{v_i,v_j},\hat{y}_{v_i,v_j}),
\label{link_loss}
\end{equation}where $\ell(\cdot,\cdot)$ is the binary cross entropy loss and $y_{v_i,v_j}\in\{0,1\}$ is the ground-truth edge existence label, where 1 indicates nodes $v_{i}$ and $v_{j}$ are connected and 0 indicates otherwise. To \jy{enable} the model \jy{to} learn  non-linkage patterns, we \jy{ augment the training set with} a set of negative edges $\mathcal{E}_{neg}$ with size $|\mathcal{E}_{tr}|$, where each element is the sampled node pair with no observed links between them in $\mathcal{G}_{tr}$, and is labeled with 0. 

\subsection{Contextualized Self-supervised Learning Task}
To enable the learned node and edge embeddings to better capture specific contextual information beneficial to link prediction, we incorporate the contextualized self-supervised learning task to predict structural context. Specifically, we use node embeddings $\bm{\Phi}(v_{i})$ and $\bm{\Phi}(v_{j})$ to \dk{respectively} predict \dk{their} structural contexts. For nodes $v_i$ and $v_j$, we respectively sample their structural contexts as $c(v_i)$ and $c(v_j)$. The existence probability of $c(v_i)$ as node $v_{i}$'s structural context,  $\hat{z}_{v_i,c(v_i)}\in[0,1]$, is modeled as
\begin{equation}
\hat{z}_{v_i,c(v_i)} = \sigma(\bm{\Phi}(v_{i})\cdot\bm{\Psi}(c(v_i))),
\label{context_vi}
\end{equation} where $\bm{\Psi}(c(v_i))$ is the embedding of structural context $c(v_i)$. Similarly, the existence probability of structural context $c(v_j)$ as node $v_{j}$'s structural context, $\hat{z}_{v_j,c(v_j)}\in[0,1]$, is modeled as
\begin{equation}
\hat{z}_{v_j,c(v_j)} = \sigma(\bm{\Phi}(v_{j})\cdot\bm{\Psi}(c(v_j))),
\label{context_vj}
\end{equation} where $\bm{\Psi}(c(v_j))$ is the embedding of structural context $c(v_j)$.

\jy{To explore the contextual information, we consider two specific strategies to sample structural contexts $c(v_i)$ and $c(v_j)$ for nodes $v_i$ and $v_j$. The details are provided below:}
\begin{itemize}
	\item \textbf{Context/neighboring nodes}. Following DeepWalk~\cite{perozzi2014deepwalk}, \jy{we consider nodes occurring in random walk context as structural context}. To construct context node embeddings, a possible solution is to perform a transformation on context node attributes similarly using Eq. (\ref{node_emb}). However, \jy{the attribute proximity among context nodes} would override the structural proximity between nodes $v_i$ and $v_j$ reflected by their common neighbors. This could be exacerbated in case of attribute noise. To retain context node identity, as a better way to construct context node embeddings, we perform a linear transformation on one-hot encodings of node IDs, which is equivalent to looking up a learnable embedding table $\bm{\Psi}\in\mathbb{R}^{|\mathcal{V}|\times d}$ with node IDs.
	
	\item \textbf{Context subgraphs}. For each node, we also consider its surrounding subgraphs as structural context, as they provide a macroscopic view to measure the structural similarity between nodes. Following~\cite{leskovec2006sampling}, we use a short random walk with fixed length starting from each node as a sample of its context subgraph. For a sampled context subgraph $\mathcal{G}_c$, its context embedding is constructed by summing the context embeddings of its nodes:
	\begin{equation*}
	\bm{\Psi}(\mathcal{G}_c) = \sum_{v\in\mathcal{G}_c}\bm{\Psi}(v),
	\end{equation*}
	where $\bm{\Psi}(v)$ is constructed via looking up a learnable embedding table $\bm{\Psi}\in\mathbb{R}^{|\mathcal{V}|\times d}$ with $v$'s node ID. 
\end{itemize}

For node pair $\langle v_i, v_j \rangle$, the contextualized self-supervised learning task aims to minimize the following loss:
\begin{equation}
\mathcal{L}_{v_i,v_j}^{context} = \ell(z_{v_i,c(v_i)},\hat{z}_{v_i,c(v_i)})+\ell(z_{v_j,c(v_j)},\hat{z}_{v_j,c(v_j)}),
\label{context_loss}
\end{equation}where $z_{v_i,c(v_i)}$ and $z_{v_j,c(v_j)}\in\{0,1\}$ are the ground-truth labels that respectively indicate whether $c(v_i)$ and ${c(v_j)}$ are nodes $v_{i}$'s and $v_{j}$'s structural contexts, with 1 for true and 0 for false. To \jy{allow} the contextualized self-supervised learning task \jy{to} learn non-context patterns, for each sampled positive structual context, we sample $k$ negative contexts with label 0. Negative context nodes are sampled as nodes not occurring in random walk context and negative context subgraphs are sampled as the sets of negative context nodes with the same node number as positive context subgraphs, which are unnecessarily connected in the training graph.

\subsection{Multi-task Learning}
To boost the link prediction task, the contextualized self-supervised learning task can be leveraged with two training strategies as follows:
\begin{itemize}
	\item \textbf{Joint Training}. \jy{We jointly train the supervised link prediction task and the self-supervised learning task} by minimizing the joint loss:
	\begin{equation}
	\mathcal{L}_{v_i,v_j}=\mathcal{L}_{v_i,v_j}^{link}+\mathcal{L}_{v_i,v_j}^{context}.
	\end{equation}
	\item \textbf{Pretraining}. We can also first train the weight matrix $\bm{W}_{emb}$ used for constructing node embeddings by minimizing the contextualized self-supervised learning loss in Eq. (\ref{context_loss}), and then finetune the model paremeters $\bm{W}_{emb}$ and $\bm{W}_{link}$ by minimizing the link prediction loss in Eq. (\ref{link_loss}). 
\end{itemize}

\begin{algorithm}[t]
	\caption{CSSL for Link Prediction (Joint Training)}
	\begin{small}
		\label{alg:CSSL:joint}
		\begin{algorithmic}[1]
			\Require A given training network $\mathcal{G}_{tr}=(\mathcal{V},\mathcal{E}_{tr},\bm{X})$;
			\Ensure Link existence probability for each unseen edge $\langle v_{i},v_{j}\rangle\in \mathcal{V}\times\mathcal{V}\setminus\mathcal{E}_{tr}$;
			\State Sample a list of positive and negative structural contexts for each node $v_i\in\mathcal{V}$;
			\State Sample a negative edge set $\mathcal{E}_{neg}$ from  $\mathcal{G}_{tr}$;
			\Repeat
			\State Sample a minibatch of edges $\{\langle v_{i_1},v_{j_1} \rangle,\cdots,\langle v_{i_n},v_{j_n} \rangle\}$
			\Statex $\quad\;$ together with their labels $\{y_{v_{i_1},v_{j_1}},\cdots, y_{v_{i_n},v_{j_n}}\}$ from 
			\Statex $\quad\;$ $\mathcal{E}_{tr}\cup \mathcal{E}_{neg}$;
			\ForEach {edge $\langle v_{i_k},v_{j_k} \rangle$ in the minibatch}
			\State Sample a $v_{i_k}$'s structural context $c(v_{i_k})$ and its label
			\Statex $\quad\;\quad\;\;$ $z_{v_{i_k},c(v_{i_k})}$ from $v_{i_k}$'s structural context list;
			\State Sample a $v_{j_k}$'s structural context $c(v_{j_k})$ and its label
			\Statex $\quad\;\quad\;\;$ $z_{v_{j_k},c(v_{j_k})}$ from $v_{j_k}$'s structural context list;
			\EndFor
			\State Update the weight matrix $\bm{W}_{link}$ by descending its 
			\Statex $\quad\;$ stochastic gradient $\nabla_{\bm{W}_{link}}\frac{1}{n}\sum_{k=1}^{n}\mathcal{L}_{v_{i_k},v_{j_k}}$;
			\State Update the context embedding matrix $\bm{\Psi}$ by descending
			\Statex $\quad\;$ its stochastic gradient $\nabla_{\bm{\Psi}}\frac{1}{n}\sum_{k=1}^{n}\mathcal{L}_{v_{i_k},v_{j_k}}$;
			\State Update the weight matrix $\bm{W}_{emb}$ by descending its 
			\Statex $\quad\;$ stochastic gradient $\nabla_{\bm{W}_{emb}}\frac{1}{n}\sum_{k=1}^{n}\mathcal{L}_{v_{i_k},v_{j_k}}$;
			\Until convergence or a fixed number of epochs expire;
			\State Infer the link existence for each unseen link $\langle v_{i},v_{j}\rangle\in {\mathcal{V}\times\mathcal{V}\setminus\mathcal{E}_{tr}}$ with the trained parameters using Eq. (\ref{link_inference}).
		\end{algorithmic}
	\end{small}
\end{algorithm}

Algorithm \ref{alg:CSSL:joint} details the training procedure of  CSSL with the joint training strategy. In Step 1, we sample a list of positive and negative structural contexts for each node $v_i\in\mathcal{V}$ with the following procedure: we first generate $\gamma$ random walks with length $l$ starting from node $v_i$; then nodes that occur following the starting node $v_i$ in random walks are collected as its positive context nodes with size $\gamma\cdot(l-1)$, or $\gamma$ positive context subgraphs are sampled for it by assembling its $l-1$ context nodes along each random walk; finally we sample $k$ negative context nodes/subgraphs for each positive context node/subgraph, and $v_i$'s context node/subgraph list is formed by combining all sampled positive and negative context nodes/subgraphs. In Step 2, the algorithm samples a set of negative edges with size $|\mathcal{E}_{tr}|$ that are not observed in $\mathcal{G}_{tr}$. Then,  the model is trained with stochastic gradient descent by sampling a minibatch of edges iteratively in Steps 3-12, where at each iteration, before performing stochastic gradient descent, for each sampled edge $\langle  v_i,v_j\rangle$, we respectively sample a context node/subgraph $c(v_i)$ and $c(c_j)$ together with the corresponding context label $z_{v_i,c(v_i)}$ and $z_{v_j,c(v_j)}$ for nodes $v_i$ and $v_j$ from their context node/subgraph lists. Finally, the existence of the unobserved links are inferred with the learned model parameters in Step 13. 


\begin{algorithm}[t]
	\caption{CSSL for Link Prediction (Pretraining)}
	\begin{small}
		\label{alg:CSSL:pretrain}
		\begin{algorithmic}[1]
			\Require A given training network $\mathcal{G}_{tr}=(\mathcal{V},\mathcal{E}_{tr},\bm{X})$;
			\Ensure Link existence probability for each unseen edge $\langle v_{i},v_{j}\rangle\in \mathcal{V}\times\mathcal{V}\setminus\mathcal{E}_{tr}$;
			\State Sample a list of positive and negative structural contexts for each node $v_i\in\mathcal{V}$;
			\State Sample a negative edge set $\mathcal{E}_{neg}$ from  $\mathcal{G}_{tr}$;
			\Repeat \Comment{model pretraining;}
			\State Sample a minibatch of edges $\{\langle v_{i_1},v_{j_1} \rangle,\cdots,\langle v_{i_n},v_{j_n} \rangle\}$
			\Statex $\quad\;$ from $\mathcal{E}_{tr}\cup \mathcal{E}_{neg}$;
			\ForEach {edge $\langle v_{i_k},v_{j_k} \rangle$ in the minibatch}
			\State Sample a $v_{i_k}$'s structural context $c(v_{i_k})$ and its label
			\Statex $\quad\;\quad\;\;$ $z_{v_{i_k},c(v_{i_k})}$ from $v_{i_k}$'s structural context list;
			\State Sample a $v_{j_k}$'s structural context $c(v_{j_k})$ and its label
			\Statex $\quad\;\quad\;\;$ $z_{v_{j_k},c(v_{j_k})}$ from $v_{j_k}$'s structural context list;
			\EndFor
			\State Update the context embedding matrix $\bm{\Psi}$ by descending
			\Statex $\quad\;$ its stochastic gradient $\nabla_{\bm{\Psi}}\frac{1}{n}\sum_{k=1}^{n}\mathcal{L}_{v_{i_k},v_{j_k}}^{context}$;
			\State Update the weight matrix $\bm{W}_{emb}$ by descending its 
			\Statex $\quad\;$ stochastic gradient $\nabla_{\bm{W}_{emb}}\frac{1}{n}\sum_{k=1}^{n}\mathcal{L}_{v_{i_k},v_{j_k}}^{context}$;
			\Until convergence or a fixed number of epochs expire;
			\Repeat \Comment{model finetuning;}
			\State Sample a minibatch of edges $\{\langle v_{i_1},v_{j_1} \rangle,\cdots,\langle v_{i_n},v_{j_n} \rangle\}$
			\Statex $\quad\;$ together with their labels $\{y_{v_{i_1},v_{j_1}},\cdots, y_{v_{i_n},v_{j_n}}\}$ from 
			\Statex $\quad\;$ $\mathcal{E}_{tr}\cup \mathcal{E}_{neg}$;
			\State Update the weight matrix $\bm{W}_{link}$ by descending its 
			\Statex $\quad\;$ stochastic gradient $\nabla_{\bm{W}_{link}}\frac{1}{n}\sum_{k=1}^{n}\mathcal{L}_{v_{i_k},v_{j_k}}^{link}$;
			\State Update the weight matrix $\bm{W}_{emb}$ by descending its 
			\Statex $\quad\;$ stochastic gradient $\nabla_{\bm{W}_{emb}}\frac{1}{n}\sum_{k=1}^{n}\mathcal{L}_{v_{i_k},v_{j_k}}^{link}$;
			\Until convergence or a fixed number of epochs expire;
			\State Infer the link existence for each unseen link $\langle v_{i},v_{j}\rangle\in {\mathcal{V}\times\mathcal{V}\setminus\mathcal{E}_{tr}}$ with the trained parameters using Eq. (3).
		\end{algorithmic}
	\end{small}
\end{algorithm}

Algorithm \ref{alg:CSSL:pretrain} details the training procedure of CSSL with the pretraining strategy. The algorithm first samples a list of positive and negative structural contexts for each node in $\mathcal{V}$ (Step 1), \jy{as well as} a set of negative edges with size $|\mathcal{E}_{tr}|$ that are not observed in $\mathcal{G}_{tr}$ (Step 2), as is done in Algorithm \ref{alg:CSSL:joint}. Then, in Steps 3-11, the model is pretrained with stochastic gradient descent by sampling a minibatch of edges and descending their context prediction loss at each iteration. After that, in Steps 12-16, the model is finetuned with the supervised link prediction task, by sampling a minibatch of edges and descending their link classification loss iteratively. Finally, in Step 17, the existence of the unobserved links \jy{is} inferred with the learned model parameters. 


\noindent\textbf{Time Complexity}. In Algorithm \ref{alg:CSSL:joint}, the time complexity \jy{of} sampling context nodes and context subgraphs is $O(|\mathcal{V}|)$. The time complexity of the training process in Steps 3-12 is $O(I|\mathcal{E}_{tr}|\bar{m}d)$, where $I$ is the number of epochs and $\bar{m}$ is the averaged number of non-zero attributes for each node. Hence, the overall time complexity of CSSL with joint training strategy is $O(|\mathcal{V}|+I|\mathcal{E}_{tr}|\bar{m}d)$. CSSL with pretraining strategy in Algorithm \ref{alg:CSSL:pretrain} shares the same time complexity. 

\section{Experiments}
In this section, we present experimental results and extensive analyses to verify the effectiveness of the proposed CSSL framework for link prediction. The experiments are conducted on a workstation with an Inter Core i9-9980XE CPU and 32G RAM, without the use of multiprocessing, GPU, or any other accelerators. 

\renewcommand\arraystretch{1.25}
\begin{table*}[t] 
	\centering
	\footnotesize
	\tabcolsep 8pt
	\caption{Summary of Benchmark Networks}
	\begin{tabular}{cccccccc} 
		\toprule
		Network & Cora & Citeseer & WebKB & Wiki & Facebook & Google+ & DBLP \\\hline
		\# of Nodes & 2,708 & 3,312 & 877 & 2,405 & 4,039 & 107,614 & 8,607 \\
		\# of Attributes & 1,433 & 3,703 & 1,703 & 4,973 & 1,406 & 19,044 & 1,102 \\
		\# of Links &5,429 & 4,715 & 1,480 & 24,357 & 88,234 & 12,238,285 & 27,614 \\
		\# of Training Links & 2,443 & 3,182 & 799 & 2,192 & 3,971 & --- & --- \\
		\# of Validation Links & 272 & 354 & 89 & 244 & 441 & --- & --- \\
		\# of Test Links & 2,714 & 1,179 & 592 & 21,921 & 83,822 & --- & --- \\
		\bottomrule
	\end{tabular}
	\label{data_summary}
\end{table*}

\renewcommand\arraystretch{1.25}
\begin{table*}[t] 
	\centering
	\tabcolsep 8pt
	\caption{Transductive Link Prediction Performance Comparison on Attributed Networks (AUC)}
	\begin{tabular}{cccccc} 
		\toprule
		Method & Cora & Citeseer & WebKB & Wiki & Facebook \\\hline
		GraphSAGE & 0.790 $\pm$ 0.010 $\bullet$ & 0.871 $\pm$ 0.009 $\bullet$ & 0.826 $\pm$ 0.024 $\bullet$ & 0.757 $\pm$ 0.019 $\bullet$ & 0.829 $\pm$ 0.016 $\bullet$\\
		Attri2Vec & 0.922 $\pm$ 0.004 $\bullet$ & 0.962 $\pm$ 0.004 $\bullet$ & 0.909 $\pm$ 0.013 $\bullet$ & 0.869 $\pm$ 0.008 $\bullet$ & 0.900 $\pm$ 0.002 $\bullet$\\
		DIG & 0.884 $\pm$ 0.003 $\bullet$ & 0.938 $\pm$ 0.006 $\bullet$ & 0.871 $\pm$ 0.022 $\bullet$ & 0.845 $\pm$ 0.014 $\bullet$ & 0.910 $\pm$ 0.005 $\bullet$\\
		GIM & 0.891 $\pm$ 0.008 $\bullet$ & 0.922 $\pm$ 0.007 $\bullet$ & 0.806 $\pm$ 0.028 $\bullet$ & 0.831 $\pm$ 0.008 $\bullet$ & 0.925 $\pm$ 0.004 $\bullet$\\
		VGAE & 0.902 $\pm$ 0.006 $\bullet$ & 0.911 $\pm$ 0.008 $\bullet$ & 0.895 $\pm$ 0.014 $\bullet$ & 0.903 $\pm$ 0.003 $\bullet$ & \textbf{0.950} $\bm{\pm}$ \textbf{0.004} $\;$\\
		SEAL & 0.739 $\pm$ 0.010 $\bullet$ & 0.805 $\pm$ 0.019 $\bullet$ & 0.812 $\pm$ 0.009 $\bullet$ & 0.810 $\pm$ 0.020 $\bullet$ & 0.881 $\pm$ 0.024 $\bullet$\\
		DEAL & 0.825 $\pm$ 0.009 $\bullet$ & 0.876 $\pm$ 0.006 $\bullet$ & 0.866 $\pm$ 0.021 $\bullet$ & 0.825 $\pm$ 0.011 $\bullet$ & 0.866 $\pm$ 0.005 $\bullet$\\
		CSSL\_Ablated & 0.910 $\pm$ 0.009 $\bullet$ & 0.953 $\pm$ 0.006 $\bullet$ & 0.931 $\pm$ 0.007 $\bullet$ & \underline{0.917 $\pm$ 0.002} $\bullet$ & 0.905 $\pm$ 0.003 $\bullet$\\
		CSSL\_Attr\_Joint & 0.916 $\pm$ 0.005 $\bullet$ & 0.965 $\pm$ 0.004 $\bullet$ & 0.940 $\pm$ 0.006 $\bullet$ & 0.916 $\pm$ 0.005 $\bullet$ & 0.915 $\pm$ 0.002 $\bullet$\\
		CSSL\_Attr\_Pretr & 0.920 $\pm$ 0.004 $\bullet$ & 0.967 $\pm$ 0.005 $\bullet$ & 0.935 $\pm$ 0.007 $\bullet$ & 0.908 $\pm$ 0.005 $\bullet$ & 0.916 $\pm$ 0.002 $\bullet$\\
		\midrule
		CSSL\_Neigh\_Joint & \textbf{0.939} $\bm{\pm}$ \textbf{0.003} $\;$ & \textbf{0.975} $\bm{\pm}$ \textbf{0.003} $\;$ & 0.947 $\pm$ 0.008 $\bullet$ & \underline{0.917 $\pm$ 0.003} $\bullet$ & 0.921 $\pm$ 0.002 $\bullet$\\
		CSSL\_Neigh\_Pretr & \underline{0.938 $\pm$ 0.004} $\;\;$ & \underline{0.974 $\pm$ 0.003} $\;\;$ & \textbf{0.953} $\bm{\pm}$ \textbf{0.004} $\;$ & 0.913 $\pm$ 0.003 $\bullet$ & \underline{0.927 $\pm$ 0.002} $\bullet$\\
		CSSL\_Subgraph\_Joint & 0.936 $\pm$ 0.005 $\bullet$ & 0.973 $\pm$ 0.003 $\bullet$ & \underline{0.950 $\pm$ 0.005} $\bullet$ & \textbf{0.921} $\bm{\pm}$ \textbf{0.003} $\;$ & 0.918 $\pm$ 0.002 $\bullet$\\
		CSSL\_Subgraph\_Pretr & 0.935 $\pm$ 0.004 $\bullet$ & 0.972 $\pm$ 0.003 $\bullet$ & 0.948 $\pm$ 0.006 $\bullet$ & 0.913 $\pm$ 0.004 $\bullet$ & 0.924 $\pm$ 0.002 $\bullet$\\
		\bottomrule
	\end{tabular}
	\label{Res_clean_network_Attribute}
\end{table*}

\subsection{Benchmark Datasets and Settings}
Seven real-world networks are used in the experiments. The detailed statistics are summarized in Table~\ref{data_summary}. 
\begin{itemize}
	\item \textbf{Cora} and \textbf{Citeseer}\footnote{https://linqs.soe.ucsc.edu/data\label{fn:linqs}} are two citation networks where nodes represent papers and links represent the citations between papers. Each paper is described by a fixed-dimension binary bag-of-words \dk{attribute} vector, \dk{with each dimension indicating the occurrence/absence of each word from a fixed-size dictionary;} \jy{the value of 1 indicates the corresponding word occurs in the paper, and 0 otherwise}.
	\item \textbf{WebKB} and \textbf{Wiki}\footref{fn:linqs} are two webpage networks where nodes indicate webpages and links indicate hyperlinks \jy{among webpages}. Node attributes are binary bag-of-words representations of webpages.
	\item \textbf{Facebook}\footnote{https://snap.stanford.edu/data/ego-Facebook.html} and \textbf{Google+}\footnote{https://snap.stanford.edu/data/ego-Gplus.html} are two social networks respectively formed by the combination of a group of Facebook and Google+ ego networks. A set of user profile values are used to construct binary node \dk{attribute vectors} according to their presence/absence.
	\item \textbf{DBLP} is the subgraph of the DBLP bibliographic network\footnote{https://aminer.org/citation (version 3)}, formed by the papers in the area of \textit{Database and Data Mining} and the citations between them. Paper titles are used to construct binary bag-of-words node \dk{attribute vectors}. The publication years of papers are also collected as their time stamps.
\end{itemize}

On \textbf{Cora}, \textbf{Citeseer}, \textbf{WebKB}, \textbf{Wiki}, and \textbf{Facebook}, we perform transductive link prediction experiments.
To ensure the training links are sparse (with roughly equal numbers of links and nodes) and prevent generating too many isolated nodes,  
we randomly split all links into training, validation and test sets, with the ratios as 45\%/5\%/50\%, 67.5\%/7.5\%/25\%, 54\%/6\%/40\%, 9\%/1\%/90\%, and 4.5\%/0.5\%/95\%, respectively. The sizes of the three sets \jy{are} given in Table~\ref{data_summary}. Following the standard setting of learning-based link prediction, for each \jy{edge}, we randomly sample one negative \jy{edge}, which is not observed in the current network. We repeat the random split ten times to avoid bias in evaluating the link prediction performance. \textbf{DBLP} is used to evaluate the performance of predicting links for out-of-sample nodes in the inductive settings, by using paper publication years to partition in-sample and out-of-sample nodes. We use \textbf{Google+} to evaluate the scalability of the proposed CSSL framework by extracting subnetworks with varying scales.

\subsection{Baseline Methods}
We compare the proposed CSSL link prediction framework with context/neighboring node prediction (\textbf{CSSL\_Neigh}) and context subgraph prediction (\textbf{CSSL\_Subgraph}) with three groups of competitive link prediction baselines:

\noindent\textbf{1) Network embedding based methods} that form edge embeddings by aggregating paired node embeddings learned via network embedding, and then convert link prediction into a link classification problem by taking edge embeddings as features. Two traditional attributed network embedding algorithms and two self-supervised based node \dk{embedding} learning algorithms are compared: 
\begin{itemize}
	\item \textbf{GraphSAGE}~\cite{hamilton2017inductive} learns node embeddings through iteratively aggregating the attributes\dk{/embeddings} of neighboring nodes.
	\item \textbf{Attri2Vec}~\cite{zhang2019attributed}  learns node embeddings by projecting node attributes into a structure-aware subspace.
	\item \textbf{DIG}~\cite{velickovic2019deep} learns node embeddings with graph convolution and self-supervision \dk{that maximizes} the mutual information between local node embeddings and the global graph \jy{representation}.
	\item \textbf{GIM}~\cite{peng2020graph} learns graph convolution based node embeddings with another self-supervision, \textit{i.e.}, maximizing the consistency between node embeddings and  the embeddings of their surrounding subgraphs. 
\end{itemize}
\textbf{2) Link modeling based methods} that directly learn a mapping from attributes and neighborhood structure of paired nodes to the link existence between them, \dk{which effectively capture the essential information in both node attributes and network structure for accurate link prediction}. 
Three state-of-the-art methods in this line are considered:
\begin{itemize}
	\item \textbf{VGAE}~\cite{kipf2016variational} reconstructs links via node embeddings obtained by graph convolution.
	\item \textbf{SEAL}~\cite{zhang2018link} learns a mapping from  extracted subgraphs \dk{including node attributes} to the link existence.
	\item \textbf{DEAL}~\cite{hao2020inductive} predicts links from aligned node attribute-oriented embeddings and structure-oriented embeddings.
\end{itemize}
\textbf{3) CSSL ablated variants} are also compared to assess the importance of different components of CSSL:
\begin{itemize}
	\item \textbf{CSSL\_Ablated} is the ablated variant without the use of self-supervised learning task.
	\item \textbf{CSSL\_Attr} is the variant that uses node attribute prediction as the self-supervised learning task. This variant is adapted from attribute prediction based pretext task designed for node classification~\cite{you2020does}.
\end{itemize}
For CSSL\_Neigh, CSSL\_Subgraph and CSSL\_Attr, we add the postfix to indicate the training strategies used: ``\_Pretr" for pretraining and ``\_Joint" for joint training.

\subsection{Experimental settings}
We use the area under the ROC curve (AUC) to evaluate the link prediction performance of all methods. For GraphSAGE, Attri2Vec and CSSL variants, we start 10 random walks with length 5 from each node. For each starting node, the nodes that follow in random walks are collected as its context nodes/subgraphs. For CSSL variants, we sample $k=1$ negative context node/subgraph for each positive context node/subgraph.
For all methods, embedding dimension is set to 128.  GraphSAGE, VGAE and the attribute-oriented embedding component of 
DEAL \jy{use} two hidden layers and the number of neurons at the first hidden layer \jy{is} set to 256. Other parameters of baselines are set to their default values. To train the proposed CSSL joint training variants, we use 100 epochs and \jy{a batch size of 20}. To train the proposed CSSL pretraining variants, we use 40 pretraining epochs, 100 finetuning epochs and \jy{a batch size of 20}. CSSL uses the Weighted-L2 as the default aggregation operator to generate edge embeddings. For VGAE, SEAL, DEAL and CSSL variants, the validation edges are used to select the best epoch. For embedding based methods (GraphSAGE, Attri2Vec, DIG and GMI), the best aggregation operators are selected from Average, Hadamard, Weighted-L1 and Weighted-L2 defined in Eq. (2)  with the validation edges.

\subsection{Transductive Link Prediction Performance Comparison}
We first compare the proposed CSSL framework with all baselines \dk{under} the transductive link prediction setting. Table~\ref{Res_clean_network_Attribute} reports the link prediction AUC scores of CSSL variants and baselines on Cora, Citeseer, WebKB, Wiki and Facebook. 
The best and the second best performers are respectively highlighted by \textbf{bold} and \underline{underline}. On each network, we also conduct paired t-test between the best performer and its competitors. The methods that are  significantly worse than the best performer at 0.05 level are marked with $\bullet$ \footnote{Tables \ref{Res_Citeseer}-\ref{Res_nonattri_network} use the same notations.}.

\begin{table*}[t] 
	\centering
	\tabcolsep 5pt
	\caption{Summary of Time Stamped Splits on DBLP}
		\begin{tabular}{ccccccccccc} 
			\toprule
			threshold year & 2000 & 2001 & 20002 & 2003 & 2004 & 2005 & 2006 & 2007 & 2008 & 2009\\\hline
			\# of In-sample Nodes & 3,979 & 4,268 & 4,632 & 4,999 & 5,401 & 5,817 & 6,307 & 6,806 & 7,359 & 7,946\\
			\# of In-sample Edges & 13,116 & 14,132 & 15,146 & 16,185 & 17,277 & 18,286 & 19,552 & 20,967 & 23,041 & 24,891\\
			\# of Out-of-sample Nodes & 4,628 & 4,339 & 3,975 & 3,608 & 3,206 & 2,790 & 2,300 & 1,801 & 1,248 & 661\\
			\# of Out-of-sample Edges & 14,498 & 13,482 & 12,468 & 11,429 & 10,337 & 9,328 & 8,062 & 6,647 & 4,573 & 2,723\\
			\bottomrule
	\end{tabular}
	\label{dblp_split_summary}
	\bigskip

	\centering
	\tabcolsep 8pt
	\caption{Out-of-sample Nodes' Link Prediction Performance Comparison on DBLP (without out-of-sample links) (AUC)}
		\begin{tabular}{ccccccccccc} 
			\toprule
			threshold year & 2000 & 2001 & 2002 & 2003 & 2004 & 2005 & 2006 & 2007 & 2008 & 2009 \\\hline
			Attri2Vec & 0.806 & 0.808 & 0.808 & 0.828 & 0.838 & 0.852 & 0.857 & 0.866 & 0.880 & 0.886\\
			DEAL & 0.756 & 0.755 & 0.761 & 0.772 & 0.782 & 0.799 & 0.790 & 0.804 & 0.816 & 0.830\\
			CSSL\_Ablated & 0.776 & 0.782 & 0.802 & 0.820 & 0.836 & 0.845 & 0.860 & 0.879 & 0.896 & 0.909\\
			CSSL\_Attr\_Joint & 0.795 & 0.802 & 0.817 & 0.833 & 0.845 & 0.871 & 0.870 & 0.889 & 0.906 & 0.916\\
			CSSL\_Attr\_Pretr & 0.786 & 0.793 & 0.802 & 0.820 & 0.842 & 0.853 & 0.858 & 0.880 & 0.894 & 0.904\\
			\midrule
			CSSL\_Neigh\_Joint & \textbf{0.811} & \textbf{0.815} & \textbf{0.825} & \textbf{0.849} & \textbf{0.857} & \textbf{0.876} & \underline{0.880} & \textbf{0.897} & 0.907 & 0.917\\
			CSSL\_Neigh\_Pretr & 0.797 & 0.812 & 0.817 & \underline{0.841} & \underline{0.853} & \underline{0.874} & \underline{0.880} & 0.895 & \textbf{0.909} & \textbf{0.921}\\
			CSSL\_Subgraph\_Joint & \underline{0.809} & \underline{0.813} & \underline{0.822} & \underline{0.841} & \textbf{0.857} & \underline{0.874} & \textbf{0.881} & \underline{0.896} & \underline{0.908} & \textbf{0.921}\\
			CSSL\_Subgraph\_Pretr & 0.799 & 0.804 & 0.814 & 0.835 & 0.850 & 0.871 & 0.873 & 0.888 & 0.906 & \underline{0.919}\\
			\bottomrule
	\end{tabular}
	\label{Res_DBLP_without_out_links}
	\bigskip

	\centering
	\tabcolsep 8pt
	\caption{Out-of-sample Nodes' Link Prediction Performance Comparison on DBLP (with out-of-sample links) (AUC)}
		\begin{tabular}{ccccccccccc} 
			\toprule
			threshold year & 2000 & 2001 & 2002 & 2003 & 2004 & 2005 & 2006 & 2007 & 2008 & 2009 \\\hline
			GraphSAGE & 0.659 & 0.597 & 0.647 & 0.647 & 0.668 & 0.651 & 0.680 & 0.656 & 0.618 & 0.651\\
			Attri2Vec & 0.803 & 0.806 & 0.809 & 0.832 & 0.838 & 0.847 & 0.855 & 0.868 & 0.874 & 0.891\\
			DIG & 0.486 & 0.529 & 0.521 & 0.519 & 0.562 & 0.565 & 0.544 & 0.491 & 0.528 & 0.559\\
			GIM & 0.747 & 0.754 & 0.770 & 0.781 & 0.801 & 0.814 & 0.827 & 0.844 & 0.859 & 0.868\\
			VGAE & 0.781 & 0.790 & 0.796 & 0.803 & 0.821 & 0.838 & 0.850 & 0.867 & 0.875 & 0.892\\
			SEAL & 0.589 & 0.574 & 0.590 & 0.596 & 0.604 & 0.604 & 0.617 & 0.639 & 0.636 & 0.642\\
			DEAL & 0.757 & 0.754 & 0.761 & 0.771 & 0.782 & 0.799 & 0.790 & 0.804 & 0.814 & 0.830\\
			CSSL\_Ablated & 0.780 & 0.790 & 0.800 & 0.819 & 0.834 & 0.856 & 0.861 & 0.878 & 0.898 & 0.911\\
			CSSL\_Attr\_Joint & 0.795 & 0.801 & 0.817 & \underline{0.833} & 0.844 & \textbf{0.870} & 0.870 & 0.890 & 0.906 & 0.916\\
			CSSL\_Attr\_Pretr & 0.787 & 0.792 & 0.802 & 0.820 & 0.841 & 0.852 & 0.856 & 0.880 & 0.893 & 0.904\\
			\midrule
			CSSL\_Neigh\_Joint & \textbf{0.808} & \textbf{0.813} & \textbf{0.825} & \textbf{0.843} & \textbf{0.860} & 0.868 & \textbf{0.880} & \textbf{0.897} & \underline{0.908} & \underline{0.920}\\
			CSSL\_Neigh\_Pretr & 0.797 & 0.808 & \underline{0.823} & \textbf{0.843} & \textbf{0.860} & \underline{0.869} & 0.876 & \textbf{0.897} & 0.904 & \underline{0.920}\\
			CSSL\_Subgraph\_Joint & \underline{0.806} & \underline{0.809} & 0.820 & \textbf{0.843} & \underline{0.856} & \underline{0.869} & \underline{0.877} & \underline{0.894} & \textbf{0.911} & \textbf{0.921}\\
			CSSL\_Subgraph\_Pretr & 0.798 & 0.801 & 0.814 & 0.832 & 0.853 & 0.863 & 0.869 & 0.891 & 0.906 & \underline{0.920}\\
			\bottomrule
	\end{tabular}
	\label{Res_DBLP_with_out_links}
\end{table*}

Table~\ref{Res_clean_network_Attribute} shows that the proposed CSSL framework achieves the best overall link prediction performance on the five networks. CSSL\_Neigh performs best on Cora and Citeseer, while the best performers on WebKB and Wiki are respectively CSSL\_Neigh\_Pretr and CSSL\_Subgraph\_Pretr.  CSSL\_Neigh\_Pretr performs slightly worse than the best performer VGAE on Facebook, while VGAE does not perform well on other four networks. The performance gain of CSSL\_Neigh and CSSL\_Subgraph over the counterpart without self-supervision (CSSL\_Ablated) and the single task based link prediction baselines proves the usefulness of structural context prediction as a self-supervised learning task for link prediction, and also the ability of CSSL to mitigate the negative impact of link sparsity via self-supervision to reinforce link prediction.


The pretraining based CSSL variants perform comparably with joint training based counterparts. Among different self-supervised learning tasks, CSSL\_Neigh and CSSL\_Subgraph perform better than CSSL\_Attr. This suggests that, as a self-supervised learning task, structural context \jy{prediction} is more effective in boosting link prediction, as compared to the attribute prediction based task.

\subsection{Inductive Link Prediction Performance Comparison}
In reality, networks are often dynamically changing with new nodes constantly joining, which may have \jy{only very few or even no connections to the existing nodes}. Therefore, we also compare the proposed CSSL framework with the baseline methods \jy{in terms of} their abilities to infer links for out-of-sample nodes on DBLP. We split the DBLP network according to a threshold on the years when papers were published. Papers published before the threshold year together with citations \jy{between these papers} form the in-sample network, and papers published after the threshold year are considered as out-of-sample nodes, with the threshold year varying from 2000 to 2009. The statistics of the splits are summarized in Table~\ref{dblp_split_summary}. We train the link prediction model on in-sample networks and then predict the links of out-of-sample nodes. Two cases are considered for the inductive link prediction setting: 1) no links are observed connecting \jy{to} out-of-sample nodes. \jy{Only Attri2Vec, DEAL and CSSL variants work in this case}, and 2) sparse (10\%) links of out-of-sample nodes are observed, including the links connecting to the in-sample training graph and the links among the out-of-sample nodes.

\begin{table*}[t] 
	\centering
	\tabcolsep 8pt
	\caption{Comparison w.r.t. Attribute Noise on Citeseer (AUC)}
		\begin{tabular}{cccccc} 
			\toprule
			noisy ratio & 5\% & 10\% & 15\% & 20\% & 25\% \\\hline
			GraphSAGE & 0.791 $\pm$ 0.010 $\bullet$ & 0.789 $\pm$ 0.012 $\bullet$ & 0.775 $\pm$ 0.010 $\bullet$ & 0.776 $\pm$ 0.015 $\bullet$ & 0.748 $\pm$ 0.009 $\bullet$\\
			Attri2Vec & 0.866 $\pm$ 0.011 $\bullet$ & 0.827 $\pm$ 0.010 $\bullet$ & 0.812 $\pm$ 0.006 $\bullet$ & 0.788 $\pm$ 0.010 $\bullet$ & 0.783 $\pm$ 0.010 $\bullet$\\
			DIG & 0.712 $\pm$ 0.015 $\bullet$ & 0.650 $\pm$ 0.025 $\bullet$ & 0.635 $\pm$ 0.011 $\bullet$ & 0.628 $\pm$ 0.009 $\bullet$ & 0.627 $\pm$ 0.011 $\bullet$\\
			GIM & 0.710 $\pm$ 0.010 $\bullet$ & 0.694 $\pm$ 0.008 $\bullet$ & 0.691 $\pm$ 0.011 $\bullet$ & 0.658 $\pm$ 0.027 $\bullet$ & 0.691 $\pm$ 0.010 $\bullet$\\
			VGAE & 0.830 $\pm$ 0.012 $\bullet$ & 0.784 $\pm$ 0.016 $\bullet$ & 0.750 $\pm$ 0.005 $\bullet$ & 0.733 $\pm$ 0.013 $\bullet$ & 0.725 $\pm$ 0.010 $\bullet$\\
			SEAL & 0.808 $\pm$ 0.012 $\bullet$ & 0.805 $\pm$ 0.008 $\bullet$ & 0.802 $\pm$ 0.014 $\bullet$ & \textbf{0.805} $\bm{\pm}$ \textbf{0.007} $\;$ & \textbf{0.801} $\bm{\pm}$ \textbf{0.011} $\;$\\
			DEAL & 0.725 $\pm$ 0.013 $\bullet$ & 0.684 $\pm$ 0.011 $\bullet$ & 0.671 $\pm$ 0.009 $\bullet$ & 0.673 $\pm$ 0.018 $\bullet$ & 0.670 $\pm$ 0.018 $\bullet$\\
			CSSL\_Ablated & 0.831 $\pm$ 0.013 $\bullet$ & 0.765 $\pm$ 0.016 $\bullet$ & 0.763 $\pm$ 0.011 $\bullet$ & 0.723 $\pm$ 0.009 $\bullet$ & 0.707 $\pm$ 0.011 $\bullet$\\
			\midrule
			CSSL\_Neigh\_Joint & \textbf{0.888} $\bm{\pm}$ \textbf{0.006} $\;$ & \textbf{0.841} $\bm{\pm}$ \textbf{0.007} $\;$ & \underline{0.836 $\pm$ 0.006} $\;\;$ & 0.797 $\pm$ 0.008 $\bullet$ & 0.784 $\pm$ 0.010 $\bullet$\\
			CSSL\_Neigh\_Pretr & 0.873 $\pm$ 0.008 $\bullet$ & 0.825 $\pm$ 0.009 $\bullet$ & 0.828 $\pm$ 0.010 $\bullet$ & 0.781 $\pm$ 0.012 $\bullet$ & 0.772 $\pm$ 0.011 $\bullet$\\
			CSSL\_Subgraph\_Joint & \underline{0.886 $\pm$ 0.007} $\;\;$ & \underline{0.840 $\pm$ 0.005} $\;\;$ & \textbf{0.840} $\bm{\pm}$ \textbf{0.005} $\;$ & \underline{0.804 $\pm$ 0.007} $\;\;$ & \underline{0.793 $\pm$ 0.006} $\;\;$\\
			CSSL\_Subgraph\_Pretr & 0.881 $\pm$ 0.008 $\bullet$ & 0.834 $\pm$ 0.010 $\;$ & 0.830 $\pm$ 0.009 $\bullet$ & 0.795 $\pm$ 0.013 $\bullet$ & 0.777 $\pm$ 0.009 $\bullet$\\
			\bottomrule
	\end{tabular}
	\label{Res_Citeseer}
\end{table*}

\begin{table*}[t]
	\centering
	\tabcolsep 8pt
	\caption{Comparison w.r.t. Attribute Noise on WebKB (AUC)}
		\begin{tabular}{cccccc} 
			\toprule
			noisy ratio & 5\% & 10\% & 15\% & 20\% & 25\% \\\hline
			GraphSAGE & 0.749 $\pm$ 0.030 $\bullet$ & 0.727 $\pm$ 0.020 $\bullet$ & 0.728 $\pm$ 0.016 $\bullet$ & 0.726 $\pm$ 0.023 $\bullet$ & 0.730 $\pm$ 0.018 $\bullet$\\
			Attri2Vec & 0.886 $\pm$ 0.008 $\bullet$ & 0.862 $\pm$ 0.011 $\bullet$ & 0.833 $\pm$ 0.010 $\bullet$ & 0.806 $\pm$ 0.016 $\bullet$ & 0.776 $\pm$ 0.038 $\bullet$\\
			DIG & 0.788 $\pm$ 0.019 $\bullet$ & 0.753 $\pm$ 0.032 $\bullet$ & 0.734 $\pm$ 0.021 $\bullet$ & 0.731 $\pm$ 0.018 $\bullet$ & 0.712 $\pm$ 0.022 $\bullet$\\
			GIM & 0.749 $\pm$ 0.025 $\bullet$ & 0.752 $\pm$ 0.021 $\bullet$ & 0.758 $\pm$ 0.008 $\bullet$ & 0.746 $\pm$ 0.024 $\bullet$ & 0.758 $\pm$ 0.007 $\bullet$\\
			VGAE & 0.793 $\pm$ 0.031 $\bullet$ & 0.754 $\pm$ 0.007 $\bullet$ & 0.713 $\pm$ 0.033 $\bullet$ & 0.558 $\pm$ 0.092 $\bullet$ & 0.501 $\pm$ 0.003 $\bullet$\\
			SEAL & 0.807 $\pm$ 0.014 $\bullet$ & 0.808 $\pm$ 0.011 $\bullet$ & 0.808 $\pm$ 0.009 $\bullet$ & 0.809 $\pm$ 0.013 $\bullet$ & 0.807 $\pm$ 0.011 $\bullet$\\
			DEAL & 0.807 $\pm$ 0.023 $\bullet$ & 0.744 $\pm$ 0.030 $\bullet$ & 0.704 $\pm$ 0.020 $\bullet$ & 0.697 $\pm$ 0.030 $\bullet$ & 0.658 $\pm$ 0.032 $\bullet$\\
			CSSL\_Ablated & 0.910 $\pm$ 0.008 $\bullet$ & 0.881 $\pm$ 0.010 $\bullet$ & 0.881 $\pm$ 0.011 $\bullet$ & 0.814 $\pm$ 0.010 $\bullet$ & 0.784 $\pm$ 0.011 $\bullet$\\
			\midrule
			CSSL\_Neigh\_Joint & 0.933 $\pm$ 0.007 $\;\;$ & 0.916 $\pm$ 0.004 $\bullet$ & 0.918 $\pm$ 0.004 $\bullet$ & 0.877 $\pm$ 0.007 $\bullet$ & 0.854 $\pm$ 0.010 $\bullet$\\
			CSSL\_Neigh\_Pretr & \textbf{0.937} $\bm{\pm}$ \textbf{0.004} $\;$ & 0.918 $\pm$ 0.007 $\;\;$ & 0.914 $\pm$ 0.004 $\bullet$ & \underline{0.887 $\pm$ 0.009} $\;\;$ & \underline{0.867 $\pm$ 0.009} $\;\;$\\
			CSSL\_Subgraph\_Joint & \underline{0.936 $\pm$ 0.005} $\;\;$ & \textbf{0.923} $\bm{\pm}$ \textbf{0.005} $\;$ & \textbf{0.921} $\bm{\pm}$ \textbf{0.005} $\;$ & 0.881 $\pm$ 0.005 $\bullet$ & 0.862 $\pm$ 0.008 $\;$\\
			CSSL\_Subgraph\_Pretr & 0.935 $\pm$ 0.005 $\;$ & \underline{0.921 $\pm$ 0.007} $\;\;$ & \underline{0.920 $\pm$ 0.004} $\;\;$ & \textbf{0.888} $\bm{\pm}$ \textbf{0.012} $\;$ & \textbf{0.870} $\bm{\pm}$ \textbf{0.014} $\;$\\
			\bottomrule
	\end{tabular}
	\label{Res_WebKB}
\end{table*}

Tables~\ref{Res_DBLP_without_out_links}-\ref{Res_DBLP_with_out_links} report the link prediction performance in the two cases, with the best and the second best performers highlighted by \textbf{bold} and \underline{underline}, respectively. From Tables~\ref{Res_DBLP_without_out_links}-\ref{Res_DBLP_with_out_links}, we can observe that the proposed CSSL variants (CSSL\_Neigh\_Joint and CSSL\_Sugbgraph\_Joint) consistently achieve the best overall performance in predicting the links \jy{for} out-of-sample nodes in both two cases, while GraphSAGE, DIG, GMI and SEAL fail to achieve satisfactory performance when out-of-sample nodes have sparse neighborhood structure. By exploiting structural context prediction based self-supervised learning, our approach can learn the mapping from node attributes to link existence with stronger generalization ability than baseline methods, thus enabling its better inductive capacity.

\subsection{Robustness against Attribute Noise}
To validate the robustness of CSSL against attribute noise, on Citeseer and WebKB, we randomly flip the binary attribute values with a ratio for each node. \jy{The rational behind is to simulate some real-world cases in social networks, where users might deliberately hide sensitive profile features or provide false profile metadata to protect their privacy, such as fabricating their gender or providing fake affiliations}. We vary the ratio from 5\% to 25\% by an increment of 5\%. 

Tables \ref{Res_Citeseer}-\ref{Res_WebKB} \jy{compare} CSSL\_Neigh and CSSL\_Subgraph with baselines in terms of different attribute noise ratios on Citeseer and WebKB, where the best and the second best performers are respectively highlighted by \textbf{bold} and \underline{underline}. As can be seen, CSSL\_Neigh and CSSL\_Subgraph consistently outperform other baselines with all attribute noise ratios on Citeseer and WebKB, except for the case \jy{with} very high noise ratios of 20\% and 25\% on Citeseer. Still for the highly noise ratio, CSSL\_Subgraph\_Joint performs comparably with the best performer SEAL that mainly leverages network structure. SEAL learns a mapping from the local neighborhood subgraph of paired nodes to the link existence with a graph neural network~\cite{zhang2018end}, where the mapping is dominated by the convolution over node structural label encoding, so the link prediction performance of SEAL
is not largely impacted by the \jy{increase in} attribute noise levels. On the other hand, 
CSSL has the capability of extracting useful information from noisy node attributes to alleviate the negative impact of noise, leading to the overall better link prediction performance at different noise ratios.

\begin{table*}[t]
	\centering
	\tabcolsep 8pt
	\tabcaption{Aggregation Operator Comparison on Cora (AUC)}
		\begin{tabular}{cccccc} 
			\toprule
			Operator & Average & Hadamard & Weighted-L1 & Weighted-L2\\\hline
			CSSL\_Neigh\_Joint & 0.655 $\pm$ 0.004 $\bullet$ & 0.910 $\pm$ 0.006 $\bullet$ & 0.930 $\pm$ 0.006 $\bullet$ & \textbf{0.939} $\bm{\pm}$ \textbf{0.003} $\;$\\
			CSSL\_Neigh\_Pretr & 0.657 $\pm$ 0.005 $\bullet$ & 0.910 $\pm$ 0.006 $\bullet$ & 0.933 $\pm$ 0.005 $\bullet$ & \textbf{0.938} $\bm{\pm}$ \textbf{0.004} $\;$\\
			CSSL\_Subgraph\_Joint & 0.655 $\pm$ 0.005 $\bullet$ & 0.902 $\pm$ 0.006 $\bullet$ & 0.929 $\pm$ 0.004 $\bullet$ & \textbf{0.936} $\bm{\pm}$ \textbf{0.005} $\;$\\
			CSSL\_Subgraph\_Pretr & 0.657 $\pm$ 0.004 $\bullet$ & 0.904 $\pm$ 0.006 $\bullet$ & 0.930 $\pm$ 0.005 $\bullet$ & \textbf{0.935} $\bm{\pm}$ \textbf{0.004} $\;$\\	
			\bottomrule
	\end{tabular}
	\label{Res_aggregator}
\end{table*}

\begin{figure*}[t]
	\centering
	\begin{minipage}[t]{0.48\textwidth}
		\centering
		\includegraphics[width=1\columnwidth]{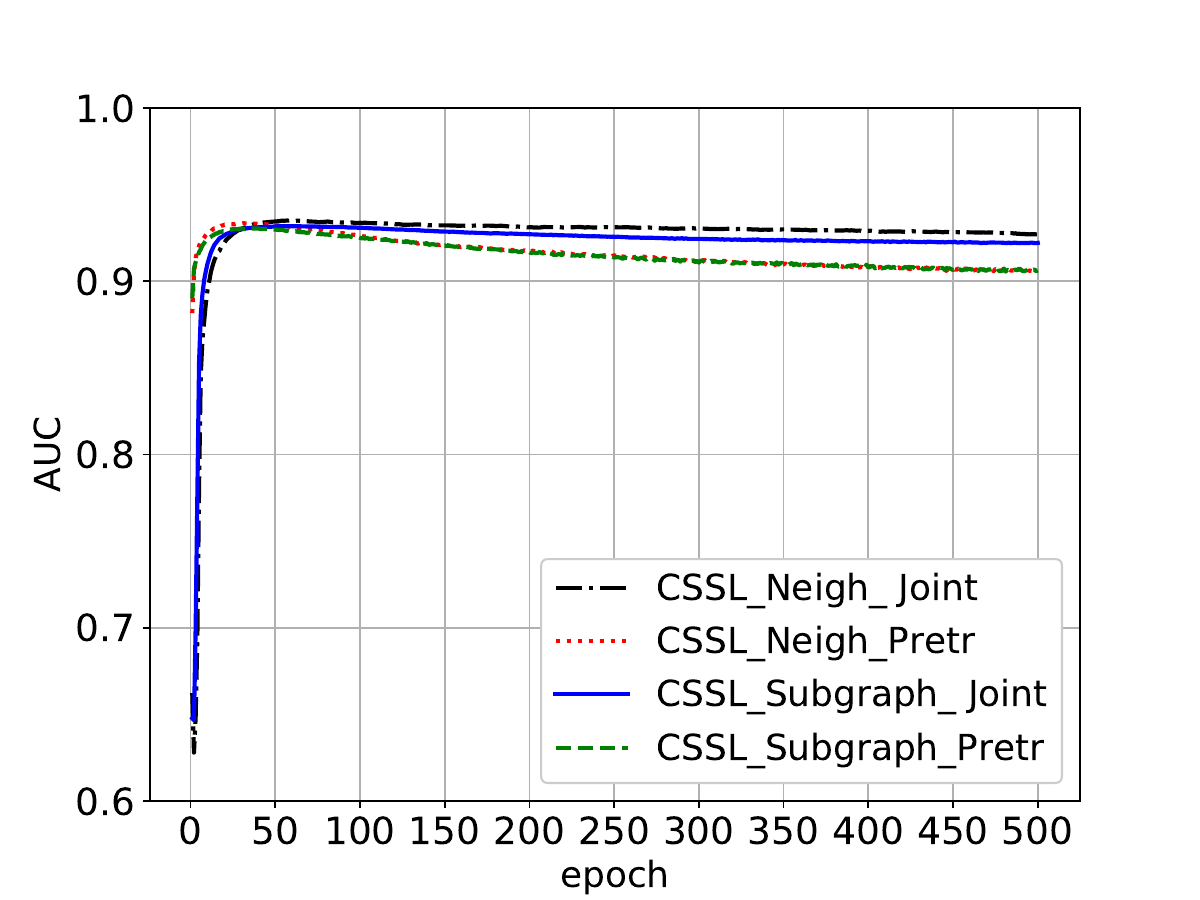}
		\caption{Convergence property of CSSL variants.}
		\label{convergence}
	\end{minipage}
	\begin{minipage}[t]{0.48\textwidth}
		\centering
		\includegraphics[width=1\columnwidth]{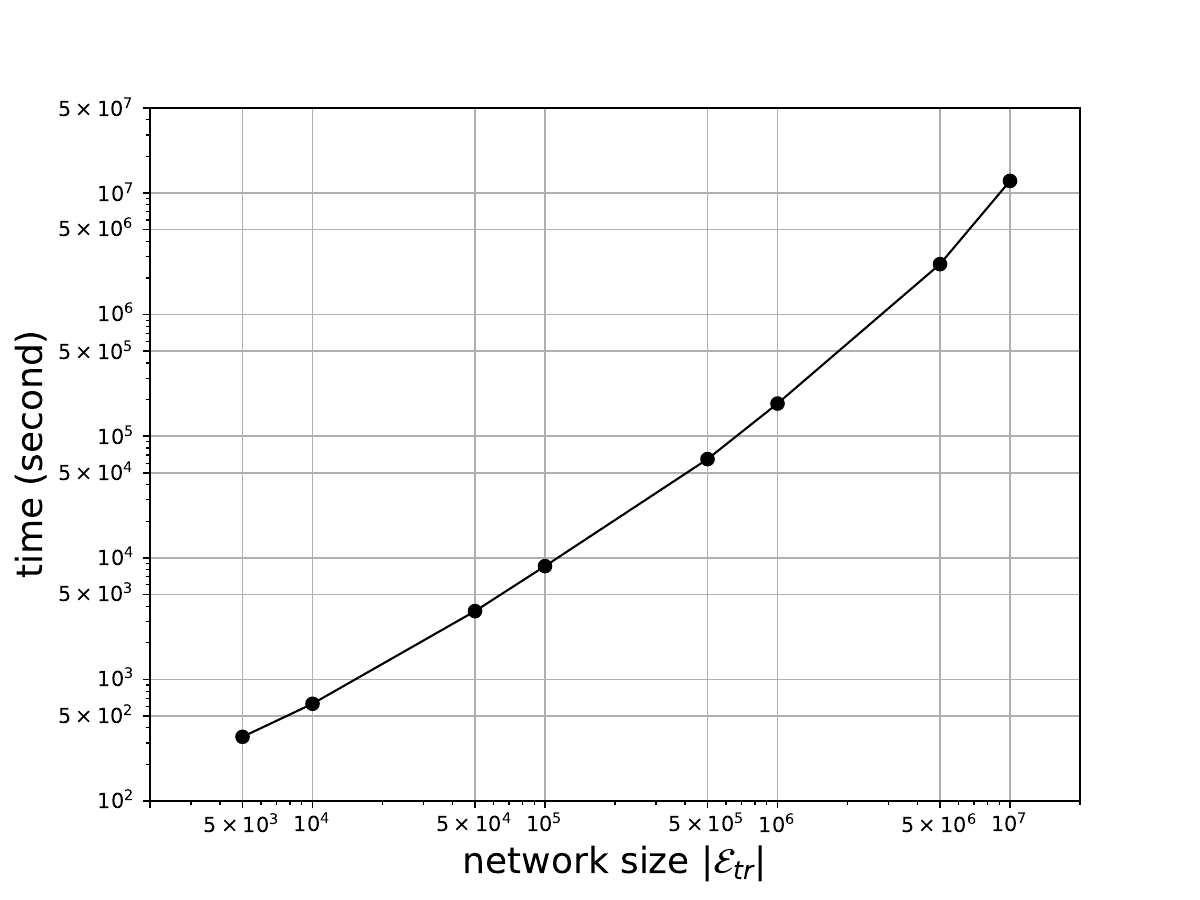}
		\caption{Scalability of CSSL\_Neigh\_Joint.}
		\label{scalblity}
	\end{minipage}
\end{figure*}

\subsection{Algorithm Convergence}
We also investigate the convergence property of the proposed CSSL framework on Cora, as a case study. Fig.~\ref{convergence} plots the link prediction performance change of all the proposed CSSL variants (CSSL\_Neigh\_Joint, CSSL\_Neigh\_Pretr, CSSL\_Subgraph\_Joint and CSSL\_Subgraph\_Pretr), when the number of epochs \jy{used for} joint training or finetuning after pretraining varies from 1 to 500. We can see that, the performance of all CSSL variants converges fast after no more than 25 epochs, and then \jy{retains} at a stable level.

\subsection{Algorithm Scalability}
We also study the scalability of the proposed CSSL with an increase of network scale, by evaluating the CPU time consumed by running CSSL\_Neigh\_Joint with 10 epochs on the sampled Google+ subnetworks of different scales.  Fig.~\ref{scalblity} plots the change of the running time as network size (the number of edges) increases from 5000 to $10^7$, with both axes in logarithm scale. As we can see, the proposed CSSL scales almost linearly with network size $|\mathcal{E}_{tr}|$.

\begin{figure*}[t]
	\centering
	\subfigure[dimension]{
		\label{fig:para:subfig:dimension} 
		\includegraphics[width=3.5in]{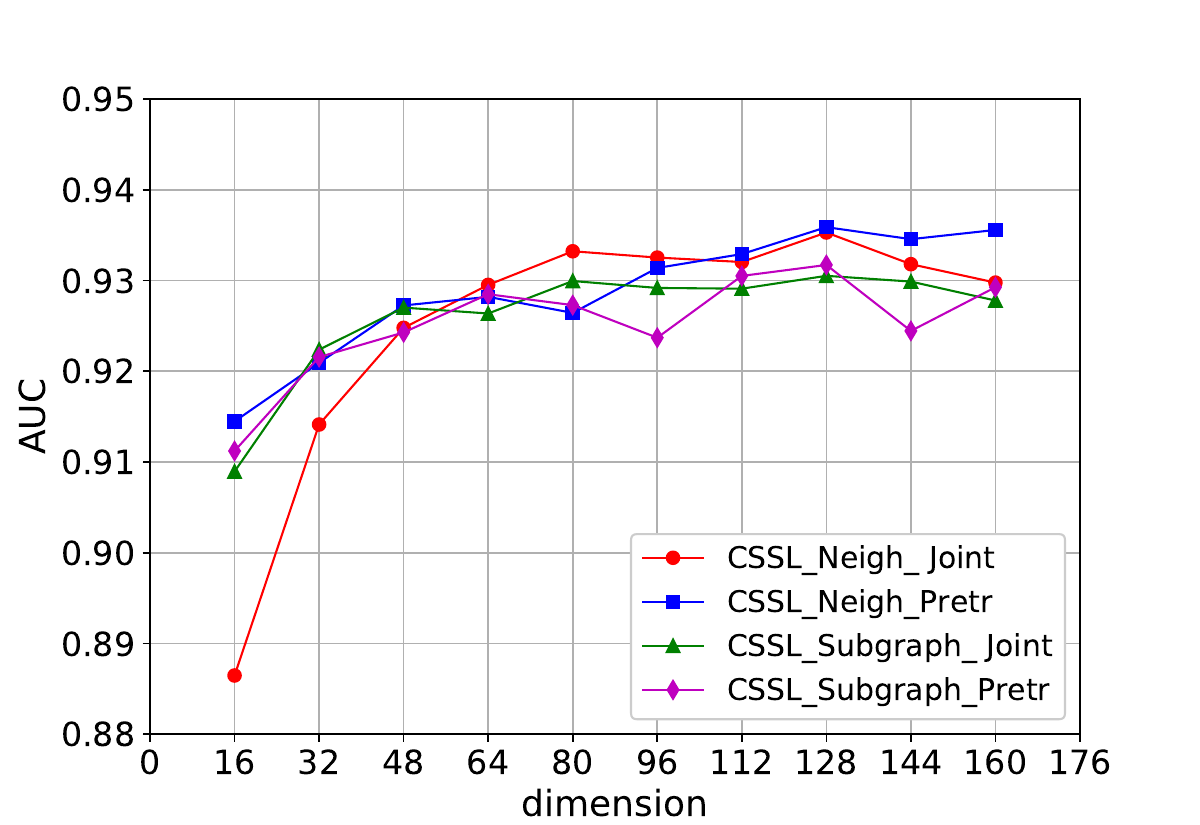}}
	\subfigure[walk length]{
		\label{fig:para:subfig:walk_length} 
		\includegraphics[width=3.5in]{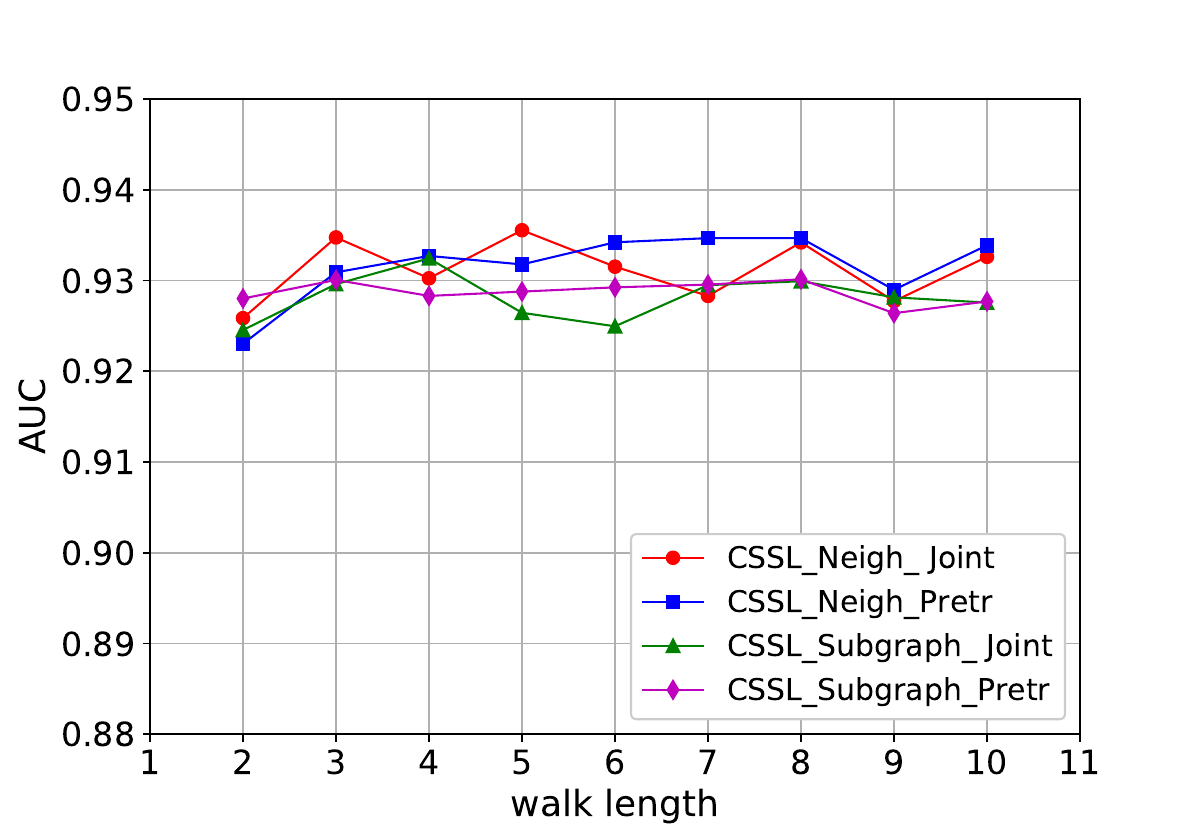}}
	\subfigure[walk number]{
		\label{fig:para:subfig:walk_number} 
		\includegraphics[width=3.5in]{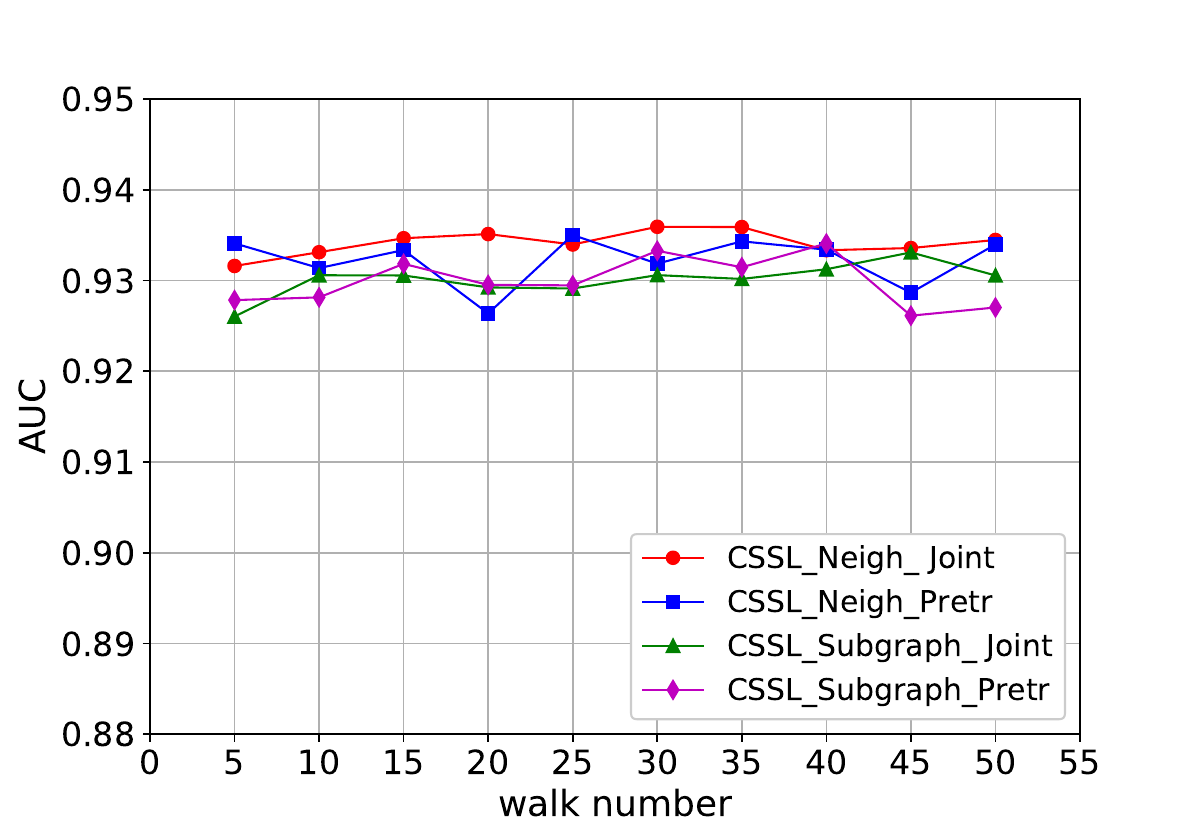}}
	\subfigure[negative sample number]{
		\label{fig:para:subfig:neg_number} 
		\includegraphics[width=3.5in]{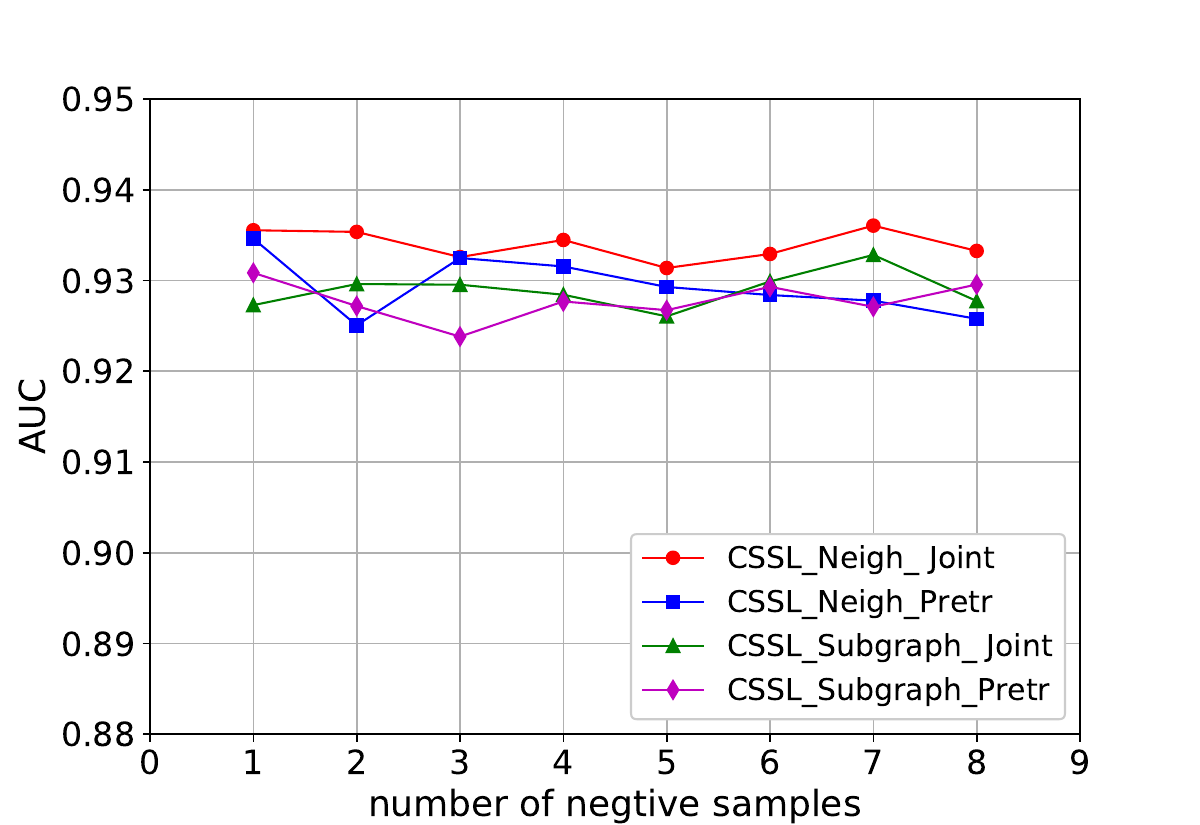}}
	\caption{Parameter sensitivity of the proposed CSSL variants in terms of (a) dimension of node/edge embeddings, (b) number of random walks starting from each node, (c) random walk length for sampling structural context, and (d) number of sampled negative context nodes/subgraphs.}
	\label{fig:para} 
\end{figure*}



\subsection{Comparison of Aggregation Operators}
As reported, the proposed CSSL uses Weighted-L2 as the aggregation operator to form edge embeddings from paired node embeddings. Now, we compare its link prediction performance \jy{using} Weighted-L2 against the other three operators, namely, Average, Hadamard and Weighted-L1. 
Table~\ref{Res_aggregator} compares \jy{the performance of CSSL variants using different aggregation operators} on Cora. For each variant, the result of the best operator is highlighted by \textbf{bold} and paired t-test is used to compare it with other operators. In general, Weighted-L1 and Weighted-L2 deliver the best link prediction performance, while Hadamard performs better than Average.  
According to the homophily phenomenon, node attribute similarity/difference is the key evidence for predicting the link existence. As Weighted-L1 and Weighted-L2 are better at characterizing node attribute difference, this translates to their superior link prediction performance.


\subsection{Parameter Sensitivity}
We also study the sensitivity of CSSL variants with respect to four important parameters, including the dimension $d$ of node/edge embeddings, the number of random walks $\gamma$ starting from each node, the random walk length $l$ for sampling structural context, and the number of sampled negative context nodes/subgraphs $k$ for each positive context. We take turns to fix any three parameters as default values and test the sensitivity of CSSL variants to the remaining one. Fig.~\ref{fig:para} shows how CSSL variants perform on Cora by varying the four parameters. Compared to other variants, CSSL\_Neigh\_Joint is more sensitive to embedding dimension. Yet, all CSSL variants tend to have stable performance with respect to varying values of these parameters.

\subsection{Generalization to Non-Attributed Networks}
So far, we have validated the efficacy of the proposed CSSL framework on attributed networks. As a flexible learning framework, CSSL is also capable of handling non-attributed networks where node attributes are unavailable. In this case, CSSL takes one-hot node representations as node attributes to infer the link existence. The baseline methods DIG, GIM, VGAE, SEAL, and DEAL use the same one-hot representation scheme to construct node features. Furthermore, we compare CSSL variants with three competitive embedding based methods that use only network structure: Node2Vec~\cite{grover2016node2vec}, the first- and second-order LINE (LINE1 and LINE2)~\cite{tang2015line}, and four heuristics-based methods with their definitions provided in Table \ref{heuristic_link_prediction}. 

\begin{table}[t]
	\centering
	\caption{Heuristics for Link Prediction between Node Pair $(v_{i},v_{j})$ with Their Direct Neighbor Sets $\mathcal{N}(v_{i})$ and $\mathcal{N}(v_{j})$}
		\begin{tabular}{lc}
			\toprule
			Heuristic & Definition\\
			\midrule
			Common Neighbors & $|\mathcal{N}(v_{i})\cap\mathcal{N}(v_{j})|$\\
			Jaccard's Coefficient & $\frac{|\mathcal{N}(v_{i})\cap\mathcal{N}(v_{j})|}{|\mathcal{N}(v_{i})\cup\mathcal{N}(v_{j})|}$\\
			Adamic-Adar Score & $\sum_{v_{k}\in\mathcal{N}(v_{i})\cap\mathcal{N}(v_{j})}\frac{1}{\log|\mathcal{N}(v_{k})|}$\\
			Preferential Attachment & $|\mathcal{N}(v_{i})|\cdot|\mathcal{N}(v_{j})|$\\
			\bottomrule
	\end{tabular}
	\label{heuristic_link_prediction}
\end{table}

\begin{table*}[t] 
	\centering
	\tabcolsep 8pt
	\caption{Comparison on Non-Attributed Networks (AUC)}
		\begin{tabular}{cccccc} 
			\toprule
			Method & Cora & Citeseer & WebKB & Wiki & Facebook\\\hline
			Common Neighbors & 0.505 $\pm$ 0.001 $\bullet$ & 0.521 $\pm$ 0.002 $\bullet$ & 0.527 $\pm$ 0.004 $\bullet$ & 0.520 $\pm$ 0.001 $\bullet$ & 0.500 $\pm$ 0.000 $\bullet$\\
			Jaccard's Coefficient & 0.505 $\pm$ 0.001 $\bullet$ & 0.521 $\pm$ 0.002 $\bullet$ & 0.527 $\pm$ 0.004 $\bullet$ & 0.520 $\pm$ 0.001 $\bullet$ & 0.500 $\pm$ 0.000 $\bullet$\\
			Adamic-Adar & 0.505 $\pm$ 0.001 $\bullet$ & 0.521 $\pm$ 0.002 $\bullet$ & 0.527 $\pm$ 0.004 $\bullet$ & 0.520 $\pm$ 0.001 $\bullet$ & 0.500 $\pm$ 0.000 $\bullet$\\
			Pref. Attachment & 0.601 $\pm$ 0.005 $\bullet$ & 0.583 $\pm$ 0.010 $\bullet$ & 0.592 $\pm$ 0.012 $\bullet$ & 0.732 $\pm$ 0.005 $\bullet$ & 0.776 $\pm$ 0.001 $\bullet$\\
			Node2Vec & 0.697 $\pm$ 0.025 $\bullet$ & 0.739 $\pm$ 0.008 $\bullet$ & 0.720 $\pm$ 0.016 $\bullet$ & 0.751 $\pm$ 0.007 $\bullet$ & 0.828 $\pm$ 0.003 $\bullet$\\
			LINE1 & 0.603 $\pm$ 0.009 $\bullet$ & 0.708 $\pm$ 0.009 $\bullet$ & 0.583 $\pm$ 0.025 $\bullet$ & 0.616 $\pm$ 0.016 $\bullet$ & 0.667 $\pm$ 0.004 $\bullet$\\
			LINE2 & 0.591 $\pm$ 0.008 $\bullet$ & 0.603 $\pm$ 0.014 $\bullet$ & 0.615 $\pm$ 0.011 $\bullet$ & 0.610 $\pm$ 0.030 $\bullet$ & 0.764 $\pm$ 0.003 $\bullet$\\
			DIG & 0.627 $\pm$ 0.007 $\bullet$ & 0.716 $\pm$ 0.015 $\bullet$ & 0.716 $\pm$ 0.015 $\bullet$ & 0.740 $\pm$ 0.010 $\bullet$ & 0.767 $\pm$ 0.003 $\bullet$\\
			GIM & 0.661 $\pm$ 0.012 $\bullet$ & 0.742 $\pm$ 0.011 $\bullet$ & 0.699 $\pm$ 0.021 $\bullet$ & 0.737 $\pm$ 0.016 $\bullet$ & 0.813 $\pm$ 0.003 $\bullet$\\
			VGAE & 0.669 $\pm$ 0.006 $\bullet$ & 0.728 $\pm$ 0.019 $\bullet$ & 0.730 $\pm$ 0.030 $\bullet$ & 0.758 $\pm$ 0.012 $\bullet$ & \underline{0.835 $\pm$ 0.006} $\bullet$\\
			SEAL & 0.772 $\pm$ 0.010 $\bullet$ & 0.836 $\pm$ 0.013 $\bullet$ & 0.841 $\pm$ 0.012 $\bullet$ & \textbf{0.795} $\bm{\pm}$ \textbf{0.002} $\;$ & 0.801 $\pm$ 0.003 $\bullet$\\
			DEAL & 0.643 $\pm$ 0.009 $\bullet$ & 0.709 $\pm$ 0.013 $\bullet$ & 0.643 $\pm$ 0.013 $\bullet$ & 0.676 $\pm$ 0.009 $\bullet$ & 0.716 $\pm$ 0.006 $\bullet$\\
			CSSL\_Ablated & 0.702 $\pm$ 0.007 $\bullet$ & 0.776 $\pm$ 0.009 $\bullet$ & 0.772 $\pm$ 0.023 $\bullet$ & 0.747 $\pm$ 0.008 $\bullet$ & 0.719 $\pm$ 0.007 $\bullet$\\
			\midrule
			CSSL\_Neigh\_Joint & 0.777 $\pm$ 0.007 $\bullet$ & \underline{0.846 $\pm$ 0.008} $\;\;$ & 0.840 $\pm$ 0.005 $\bullet$ & 0.768 $\pm$ 0.009 $\bullet$ & 0.801 $\pm$ 0.005 $\bullet$\\
			CSSL\_Neigh\_Pretr & \underline{0.779 $\pm$ 0.009} $\;\;$ & 0.845 $\pm$ 0.007 $\bullet$ & 0.850 $\pm$ 0.006 $\bullet$ & 0.780 $\pm$ 0.009 $\bullet$ & 0.823 $\pm$ 0.005 $\bullet$\\
			CSSL\_Subgraph\_Joint & 0.774 $\pm$ 0.008 $\bullet$ & 0.841 $\pm$ 0.006 $\bullet$ & \underline{0.854 $\pm$ 0.009} $\bullet$ & 0.786 $\pm$ 0.005 $\bullet$ & 0.826 $\pm$ 0.004 $\bullet$\\
			CSSL\_Subgraph\_Pretr & \textbf{0.780} $\bm{\pm}$ \textbf{0.007} $\;$ & \textbf{0.851} $\bm{\pm}$ \textbf{0.005} $\;$ & \textbf{0.862} $\bm{\pm}$ \textbf{0.006} $\;$ & \underline{0.788 $\pm$ 0.008} $\bullet$ & \textbf{0.846} $\bm{\pm}$ \textbf{0.005} $\;$\\
			\bottomrule
	\end{tabular}
	\label{Res_nonattri_network}
\end{table*}

Table~\ref{Res_nonattri_network} reports the link prediction results on five networks without node attributes, with the best and the second best performers highlighted by \textbf{bold} and \underline{underline}, respectively. Clearly, as compared to Table~\ref{Res_clean_network_Attribute}, significant performance drops can be observed when node attributes are not incorporated for all methods except for SEAL. As SEAL does not well leverage node attributes for link prediction, it even achieves better performance on Cora, Citeseer, WebKB and Wiki \jy{at the absence of} node attributes.

To fully exploit network structure, for CSSL\_Neigh and CSSL\_Subgraph, we respectively set the number of random walks starting from per node and the random walk length to 80 and 20, and we use Weighted-L1 aggregation operator to form edge embeddings for all CSSL variants. For Node2Vec, we start 10 random walks with length 80 from each node and collect context nodes with window size 10, with the default random walk setting $p=q=1$ used. Default parameters are used to train LINE.

Note that even at the absence of node attributes, CSSL\_Neigh and CSSL\_Subgraph still markedly outperform other baselines, with the CSSL\_Subgraph\_Pretr variant yielding the best overall performance. Specially, CSSL\_Neigh and CSSL\_Subgraph outperform CSSL\_Ablated without self-supervision by a large margin. Again, this validates the effectiveness of structural context prediction in boosting link prediction, as a self-supervised learning task, especially for the case without node attributes. 

Apart from CSSL variants, SEAL generally achieves the second best performance, \jy{far} superior to other rigid heuristics-based methods. This proves the efficacy of SEAL in extracting useful node neighborhood structure to boost the link prediction performance. Among network embedding based methods, DeepWalk achieves the best performance, which attributes to its special ability in exploring high-order neighborhood structure. As a whole, traditional heuristics-based methods do not exhibit competitive results as compared to other advanced baselines.

On non-attributed networks, the pretraining based CSSL variants overall perform better than the joint training based counterparts, because the pretraining strategy helps better learn informative node structure embeddings, which are essential for link prediction. In comparison, on attributed networks, with the use of informative node attributes to construct node embeddings, the pretraining and joint training variants perform comparably.

\section{Conclusion}
In this paper, we proposed a novel contextualized self-supervised learning framework for link prediction, which is \dk{achieved} by predicting the existence of structural context. The proposed CSSL framework is generic, with the ability to \dk{leverage} various types of structural context, the capacity to predict links for both in-sample and out-of-sample nodes, and the flexibility to handle both attributed and non-attributed networks. The proposed link prediction framework scales linearly with the number of edges, which endows it with the potential to handle large-scale networks. Through the proposed self-supervised learning task, the CSSL framework makes the best of useful information in sparse links and noisy node attributes to perform link prediction. Comprehensive experiments on seven real-world networks demonstrated that the proposed CSSL framework achieves the state-of-the-art link prediction performance, on the transductive/inductive, attributed/non-attributed, and attribute-clean/attribute-noisy settings.

\section*{Acknowledgments}
This work is supported by a joint CRP research fund between the University of Sydney and Data61, CSIRO, and in part by NSF under grants III-1763325, III-1909323, III-2106758, and SaTC-1930941. 
\ifCLASSOPTIONcaptionsoff
  \newpage
\fi



%
\bibliographystyle{IEEEtran}
\bibliography{cssl-tkde.bib}

\vspace{-0.5cm}
\begin{IEEEbiography}[{\includegraphics[width=0.9in,clip]{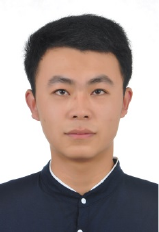}}]{Daokun Zhang}
	received the Ph.D. degree in computer science from the University of Technology Sydney (UTS), Australia, in 2019. He is currently a Research Fellow at the Department of Data Science \& AI, Faculty of Information Technology, Monash University, Australia. His research interests include graph machine learning and its applications. 
\end{IEEEbiography}

\vspace{-0.5cm}
\begin{IEEEbiography}[{\includegraphics[width=0.9in,clip]{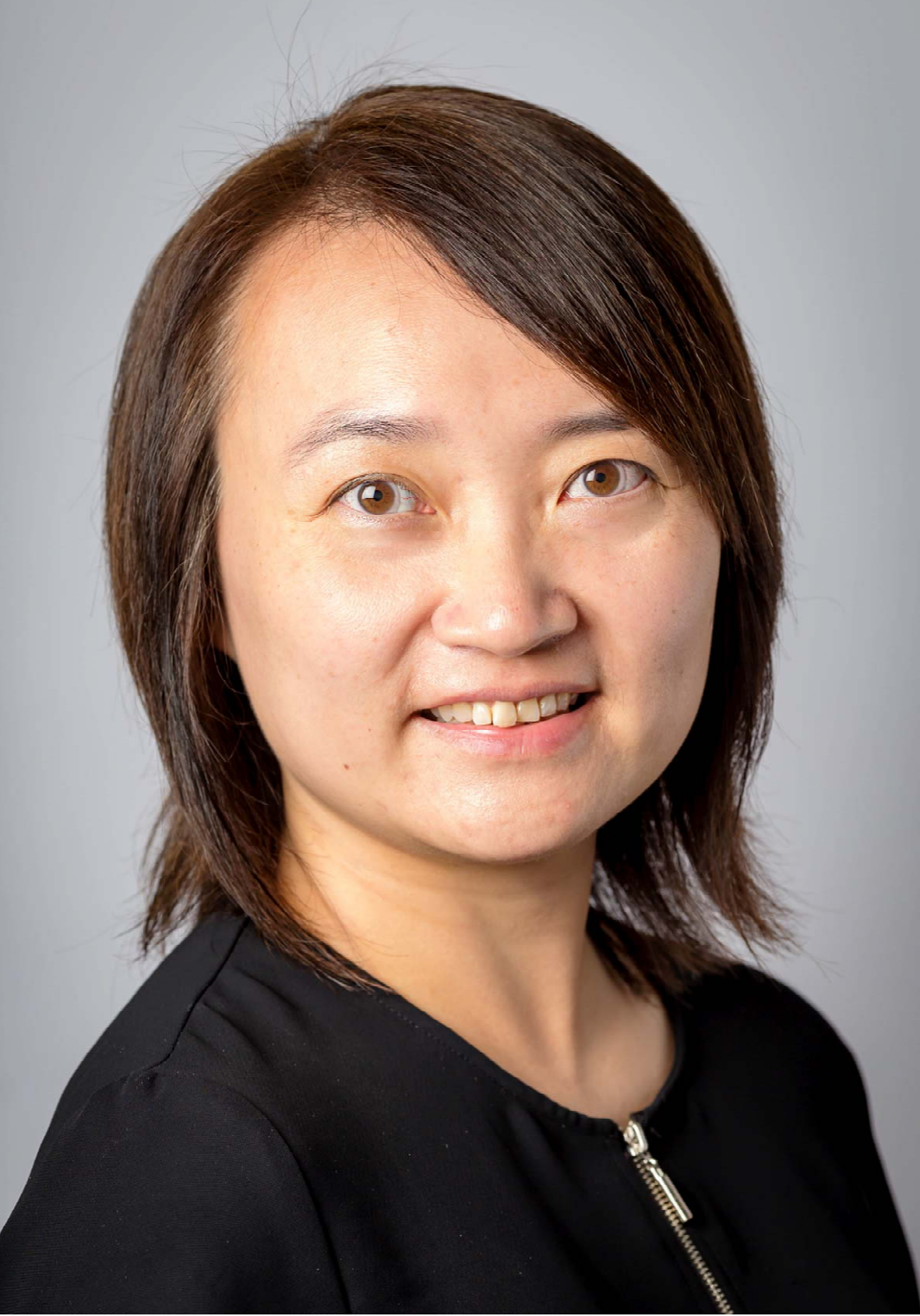}}]{Jie Yin}
	received the PhD degree in Computer Science from the Hong Kong University of Science and Technology, Hong Kong. She is currently an Associate Professor at the Discipline of Business Analytics, The University of Sydney, Australia. Her research interests include data mining, machine learning, and interpretable AI. She has published more than 70 refereed journal and conference papers in these areas. She is a co-chair of the International Workshop on Social Web for Disaster Management (SWDM 2015--2018). She is an Associate Editor of IEEE Transactions on Big Data.
\end{IEEEbiography}

\vspace{-0.5cm}
\begin{IEEEbiography}[{\includegraphics[width=0.9in,clip]{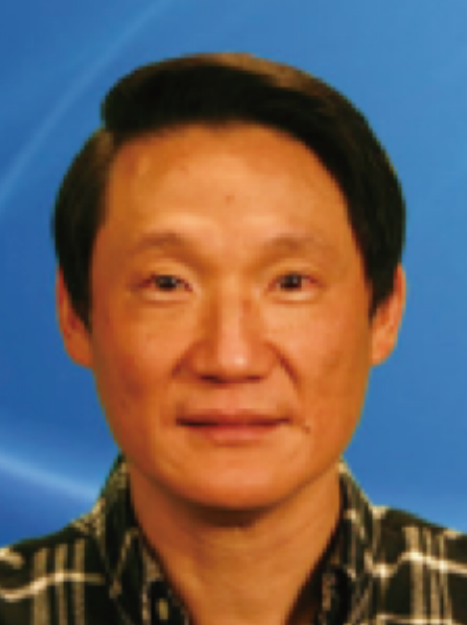}}]{Philip S. Yu}
	(Life Fellow, IEEE) received the Ph.D. degree in electrical engineering from Stanford University, CA, USA. He is a Distinguished Professor of computer science with the University of Illinois at Chicago, Chicago, IL, USA, where he is also the Wexler Chair in Information Technology. He has published more than 800 articles in refereed journals and conferences. He holds or has applied for more than 300 U.S. patents. His research interests include big data, data mining, data streams, databases, and privacy.
	Dr. Yu is a fellow of the ACM. He received the ACM SIGKDD 2016 Innovation Award, the Research Contributions Award from the IEEE International Conference on Data Mining in 2003, and the Technical Achievement Award from the IEEE Computer Society in 2013. 
\end{IEEEbiography}

\end{document}